\documentclass[10pt,a4paper,twocolumn]{article}
\usepackage[margin=1.75cm,columnsep=0.75cm]{geometry}
\usepackage{amsmath,amssymb,bm,mathtools}
\usepackage{booktabs,array,multirow}
\usepackage{graphicx}
\usepackage{xcolor}
\usepackage{hyperref}
\usepackage{microtype}
\usepackage{caption}
\usepackage{subcaption}
\usepackage{siunitx}
\usepackage{placeins}
\usepackage{float}
\usepackage{tikz-feynman}
\usepackage[T1]{fontenc}
\usepackage{lmodern}
\usepackage{dblfloatfix}
\usepackage{cuted}
\usepackage{balance}
\usepackage{orcidlink}
\usepackage{flafter}

\hypersetup{colorlinks=true,linkcolor=blue!55!black,citecolor=blue!55!black,urlcolor=blue!55!black}
\allowdisplaybreaks
\newcommand{\ttbar}{t\bar t}
\newcommand{\mtt}{m_{t\bar t}}
\newcommand{\Lr}{\Lambda_r}
\newcommand{\mph}{m_\phi}
\newcommand{\fshape}{f_{\rm shape}}
\newcommand{\CLs}{CL_s}
\newcommand{\shat}{\hat s}
\newcommand{\as}{\alpha_s}

\begin{document}
\twocolumn[
\begin{@twocolumnfalse}
\begin{center}
{\LARGE\bfseries Radion--QCD Interference in $t\bar t$ Production at the HL-LHC:\\[0.2em]
Finite-Top-Mass Effects and Projected Sensitivity\par}
\vspace{1.0em}
{\large A. Bellagroudi\orcidlink{0009-0002-7563-1176}$^{1,*}$
, F. Fassi\orcidlink{0000-0002-6423-7213}$^{1}$\par}
\vspace{0.45em}
{$^{1}$Laboratory of Condensed Matter and Interdisciplinary Sciences, Unit\'e de Recherche Labellis\'ee CNRST (URL-CNRST), Faculty of Sciences, Mohammed V University in Rabat, Rabat 1014, Morocco. \par}

\vspace{0.35em}
{\small $^{*}$Corresponding author: \texttt{ahmed.bellagroudi@cern.ch}\\
F. Fassi: \texttt{farida.fassi@cern.ch}
\par}
\end{center}
\vspace{0.7em}
\begin{abstract}
We investigate the interference of a heavy radion with the QCD continuum
in top-quark-pair production at the High-Luminosity Large Hadron Collider,
restricting the numerical study to the pure-radion limit $\xi=0$.
The gluon-fusion amplitude combines the QCD trace anomaly with the exact
finite-top-mass loop form factor, whose coherent sum fixes both the
magnitude and the phase of the production coefficient. Above the top
threshold the loop develops an absorptive part, rendering the production
coefficient complex and allowing the interference to remain non-zero on
the resonance pole, where a purely real point-like coefficient would give
none. The complex coefficient is validated independently using a native
loop-induced implementation and a common-event phase-basis construction.
The resulting signature is a peak--dip deformation rather than a positive
bump, while detector smearing turns the narrow truth-level structure into
a broad sub-percent distortion.

For $3\,\mathrm{ab}^{-1}$ at $\sqrt{s}=14$~TeV, we construct an
ATLAS-anchored phenomenological response and perform an exact binned
Poisson Asimov analysis with a free background normalization and a
correlated shape nuisance. For a $0.1\%$ shape benchmark with a
$50$~GeV correlation length, the profiled median 95\% $CL_s$ reach in
$\Lambda_r$ is $2.49$, $2.82$, $3.00$, $2.56$, and $2.19$~TeV for
radion masses of $600$, $800$, $1000$, $1200$, and $1500$~GeV,
respectively. A diagnostic decomposition at $800$~GeV gives
$2.18$~TeV for the resonance-squared term alone and $2.83$~TeV for
the interference term alone. A selected-background deformation stress
test largely removes the apparent maximum near $1$~TeV, showing that
its location and prominence are normalization-prescription dependent,
while leaving the central conclusion unchanged: the projected sensitivity
is predominantly interference driven.
\end{abstract}
\vspace{0.4em}
\vspace{1.0em}
\end{@twocolumnfalse}
]
\section{Introduction}
Warped extra dimensions provide a geometric mechanism for separating the electroweak and gravitational scales. In the two-brane Randall--Sundrum (RS) construction~\cite{RS1999}, the warp factor exponentially redshifts scales between the ultraviolet and infrared branes. Stabilization of the interbrane separation by a bulk scalar field in the Goldberger--Wise mechanism~\cite{GW1999} leaves a physical scalar modulus, the radion, whose interactions are governed by the trace of the Standard Model (SM) energy--momentum tensor~\cite{Csaki2001,Bae2000,Dominici2003}.

Recent radion collider phenomenology has also revisited anomaly-induced
gauge couplings and Higgs--radion mixing in complementary collider
settings~\cite{Cox2014,AitTamlihat2026}.

A heavy neutral spin-zero state produced through gluon fusion and decaying to $\ttbar$ does not generically appear as a simple Breit--Wigner excess. The resonance amplitude interferes coherently with the large SM QCD continuum and can generate peak--dip, dip--peak, or deficit-like structures. This is well established for generic heavy scalars and pseudoscalars~\cite{Hespel2016,CarenaLiu2016} and is incorporated in dedicated ATLAS and CMS interference-aware $\ttbar$ searches~\cite{ATLAS2024,CMS2025}. A broader recent CMS resonance search combining the all-hadronic, single-lepton, and dilepton channels provides complementary Run-2 context, including spin-zero interpretations over part of the mass range~\cite{CMSB2G2026}.

The radion has an additional model-specific ingredient: its coupling to gluons is the coherent sum of a QCD trace-anomaly contact term and the resolved heavy-quark loop. Above the $\ttbar$ threshold the top loop develops an absorptive part, so the physical production coefficient becomes complex. This phase is especially important for interference because it permits a non-zero on-pole contribution that is absent for a purely real point-like scalar coefficient.

The goal of this study is to isolate and validate this radion-specific interference structure and quantify its impact on HL-LHC sensitivity after realistic reconstructed-mass smearing and nuisance profiling. The numerical analysis is deliberately restricted to the pure-radion limit $\xi=0$; a full Higgs--radion mixing scan would require recomputing the physical couplings, widths, production phase, branching fractions, and constraints and is left for future work.

\section{Warped radion framework}
\subsection{Five-dimensional origin}
The RS geometry is a slice of $\mathrm{AdS}_5$,
\begin{equation}
 ds^2=e^{-2kr_c|\varphi|}\eta_{\mu\nu}dx^\mu dx^\nu-r_c^2d\varphi^2,
 \qquad \varphi\in[-\pi,\pi],
\end{equation}
with curvature $k$, compactification radius $r_c$, and the infrared brane at $\varphi=\pi$. Goldberger--Wise stabilization fixes the interbrane separation and generates a non-zero radion mass~\cite{GW1999}. After dimensional reduction and canonical normalization, the radion interaction scale is
\begin{equation}
 \Lr\simeq \sqrt6\,\bar M_{\rm Pl}e^{-kL},
\end{equation}
where $L$ denotes the stabilized proper separation.

\subsection{Effective action and mixing convention}
A convenient low-energy description is
\begin{equation}
\begin{aligned}
S_{\rm eff}=\int d^4x\Big[&\mathcal L_{\rm SM}+\frac12(\partial r_0)^2
-\frac12m_{r_0}^2r_0^2\\
&+\frac{r_0}{\Lr}T^\mu_{\ \mu}\Big]+S_\xi,
\end{aligned}
\end{equation}
with the curvature--Higgs operator
\begin{equation}
S_\xi=-\xi\int d^4x\sqrt{-g_{\rm vis}}\,R(g_{\rm vis})H^\dagger H.
\end{equation}
Defining
\begin{equation}
\gamma=\frac{v}{\Lr},\qquad Z^2=1+6\xi\gamma^2(1-6\xi),
\end{equation}
and writing the gauge eigenstates as
\begin{equation}
r_0=A\phi+Bh,\qquad h_0=C\phi+Dh,
\end{equation}
the convention used here is~\cite{Dominici2003,Cox2014}
\begin{align}
 A&=-\frac{\cos\theta}{Z}, & B&=\frac{\sin\theta}{Z},\\
 C&=\sin\theta+\frac{6\xi\gamma}{Z}\cos\theta, & D&=\cos\theta-\frac{6\xi\gamma}{Z}\sin\theta.
\end{align}
The normalized top coupling of the heavy physical state is
\begin{equation}
\kappa_t^\phi=C+\gamma A.
\end{equation}

\subsection{Pure-radion limit}
For the numerical analysis $\xi=0$, hence $Z=1$, $\theta=0$, and
\begin{equation}
A=-1,\qquad B=0,\qquad C=0,\qquad D=1,
\end{equation}
so that
\begin{equation}
\kappa_t^\phi=-\gamma=-\frac{v}{\Lr}.
\end{equation}
With the sign convention used in the generator mapping,
\begin{equation}
\boxed{\mathcal L_{\phi t\bar t}=+\frac{m_t}{\Lr}\phi\bar t t.}
\end{equation}
The same real radion-top coupling enters the resolved production loop and the $\phi\to\ttbar$ decay amplitude, producing the expected overall $\Lr^{-2}$ scaling of the resonance amplitude. The non-trivial complex phase instead originates from the absorptive part of the production loop above threshold.

\subsection{Widths}
At $\xi=0$ the width model includes $WW$, $ZZ$, $t\bar t$, $hh$, and $gg$. The explicit pure-radion partial-width formulae are collected in Appendix~\ref{app:purewidths}; the gluonic channel uses the same anomaly-plus-resolved-top convention as the production amplitude. The benchmark total widths at $\Lr=3$ TeV are listed in Table~\ref{tab:widths}.

\begin{figure}[htbp]
\centering
\includegraphics[width=\columnwidth]{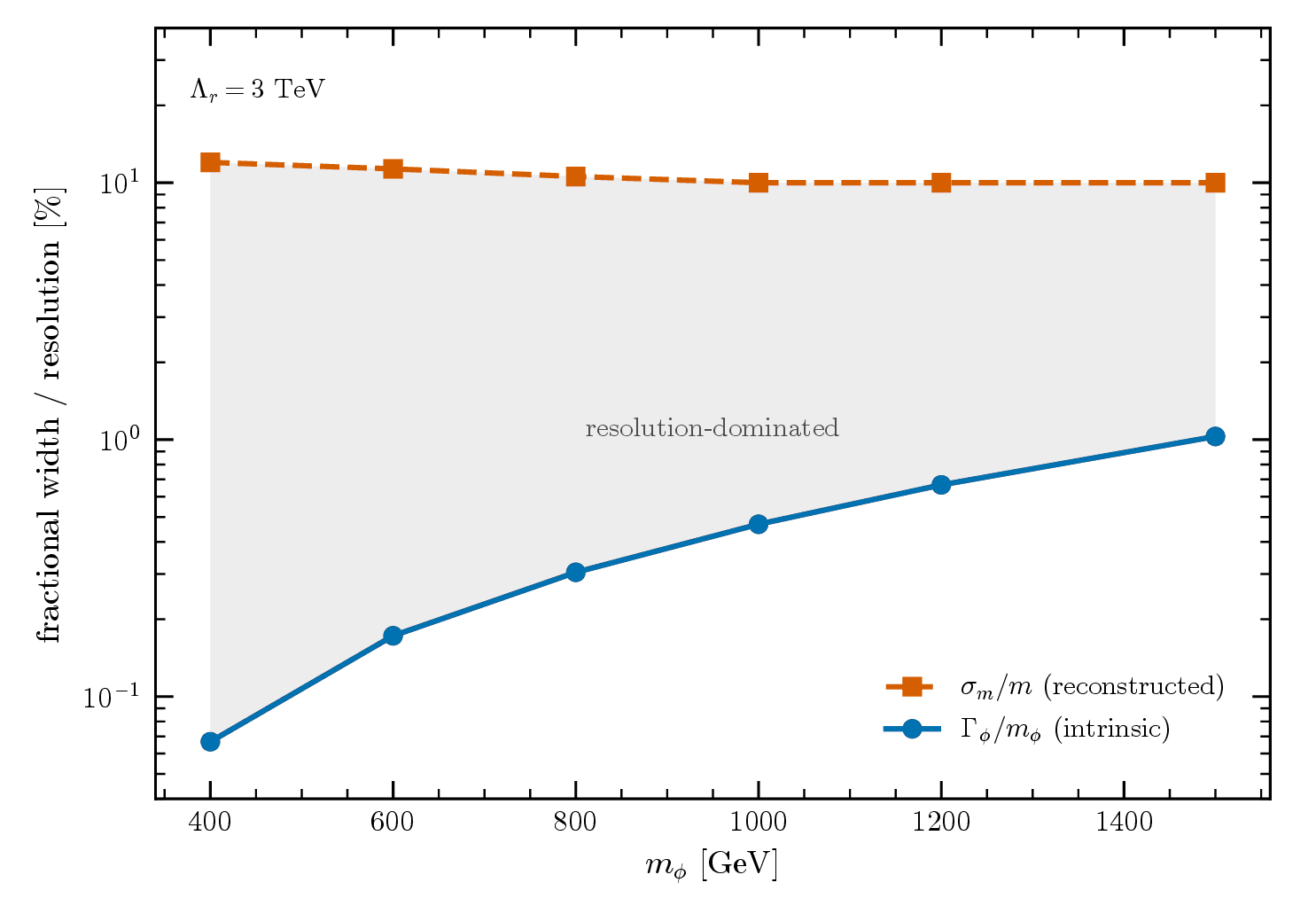}
\caption{Intrinsic radion width compared with the nominal reconstructed-$\mtt$ resolution at $\Lr=3$ TeV. The detector response is broader than the intrinsic width throughout the benchmark range, so the observed morphology is detector-resolution dominated.}
\label{fig:widthres}
\end{figure}

\begin{table}[htbp]
\centering
\caption{Pure-radion benchmark widths at $\Lr=3$ TeV. The independent width-model closure gives $\Gamma_\phi(800)=2.436$ GeV, consistent with the 2.44 GeV event benchmark.}
\label{tab:widths}
\begin{tabular}{rcc}
\toprule
$\mph$ [GeV] & $\Gamma_\phi$ [GeV] & $\Gamma_\phi/\mph$ \\
\midrule
400 & 0.267 & 0.0667\% \\
600 & 1.035 & 0.1724\% \\
800 & 2.440 & 0.3050\% \\
1000 & 4.686 & 0.4686\% \\
1200 & 8.000 & 0.6666\% \\
1500 & 15.438 & 1.0292\% \\
\bottomrule
\end{tabular}
\end{table}
Because every coupling scales as $\Lr^{-1}$,
\begin{equation}
\boxed{\Gamma_\phi(\Lr)=\Gamma_0\left(\frac{\Lambda_{r,0}}{\Lr}\right)^2.}
\label{eq:widthscaling}
\end{equation}
The intrinsic state is substantially narrower than the reconstructed $\mtt$ resolution throughout the benchmark grid, as shown in Fig.~\ref{fig:widthres}.

\section{Loop-induced gluon fusion and complex production phase}
\subsection{Trace anomaly and resolved top loop}
The radion couples to the trace of the SM energy--momentum tensor,
\begin{equation}
T^\mu_{\ \mu}=T^\mu_{\ \mu}\big|_{\rm classical}+T^\mu_{\ \mu}\big|_{\rm anomaly}.
\end{equation}
The classical part contains the explicit mass terms, whereas the gluon channel receives the QCD scale anomaly,
\begin{equation}
T^\mu_{\ \mu}\big|_{\rm anomaly}\supset \frac{\beta_s(g_s)}{2g_s}G^a_{\mu\nu}G^{a\mu\nu}.
\end{equation}
In the convention used throughout the calculation, the anomaly contact contribution is represented by
\begin{equation}
b_6=11-\frac23n_f=7\qquad(n_f=6),
\end{equation}
while the finite-mass top triangle is retained explicitly.

The connection with heavy-quark decoupling can be made explicit from the one-loop QCD running. Defining the dimensionless coefficient $\beta_0^{(n_f)}$ through
\begin{equation}
\mu^2\frac{d\alpha_s}{d\mu^2}
=-\frac{\beta_0^{(n_f)}}{4\pi}\alpha_s^2+\mathcal O(\alpha_s^3),
\qquad
\beta_0^{(n_f)}=11-\frac23n_f,
\end{equation}
continuity of $\alpha_s$ across a heavy-quark threshold $M$ gives the one-loop matching relation
\begin{equation}
\Lambda_{n_f-1}
=\Lambda_{n_f}^{\frac{33-2n_f}{35-2n_f}}
M^{\frac{2}{35-2n_f}}.
\end{equation}
At the top threshold, $n_f=6$ and $M=m_t$, so
\begin{equation}
\Lambda_5=\Lambda_6^{21/23}m_t^{2/23},
\qquad
\beta_0^{(6)}=7,
\qquad
\beta_0^{(5)}=\frac{23}{3}.
\end{equation}
Consequently,
\begin{equation}
\beta_0^{(5)}-\beta_0^{(6)}=\frac23.
\end{equation}
This is exactly the contribution recovered from the formal heavy-top limit of the resolved triangle,
\begin{equation}
\frac12A_{1/2}\xrightarrow[m_t^2\gg\shat]{}\frac12\left(\frac43\right)=\frac23.
\end{equation}
Thus the anomaly-plus-resolved-top prescription has the expected decoupling bookkeeping: the six-flavour contact coefficient $7$ plus the heavy-top loop tends to the five-flavour coefficient $23/3$. This threshold relation is used only as a theoretical consistency check; the numerical analysis retains the full finite-top-mass form factor rather than integrating out the top quark.

Matching the contact and loop pieces in a common generator normalization therefore gives
\begin{equation}
\boxed{F_\phi(\shat)=7+\frac12A_{1/2}(\shat).}
\end{equation}
The heavy-top limit $A_{1/2}\to4/3$ then gives $F_\phi\to23/3$, providing an independent normalization and decoupling check. The value $23/3$ must not be used simultaneously as the contact term when the explicit full top loop is retained, since that would double count the heavy-top contribution.

\subsection{Finite-top-mass form factor}
\begin{figure*}[tbp]
\centering
\includegraphics[width=0.96\textwidth]{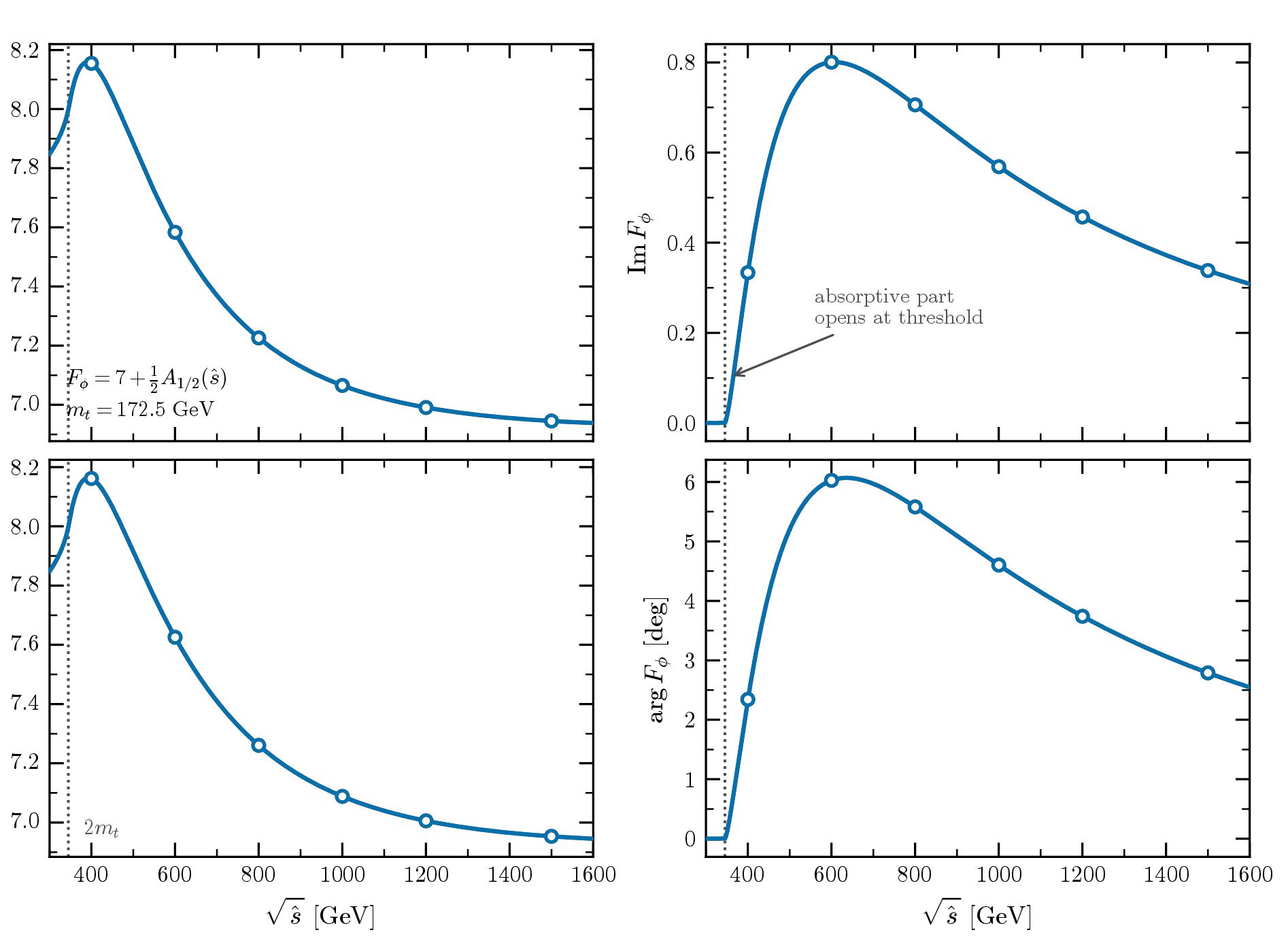}
\caption{Resolved radion production form factor $F_\phi(\shat)=7+\tfrac12A_{1/2}(\shat)$ for $m_t=172.5$ GeV. The absorptive part vanishes below $2m_t$ and turns on above threshold, giving the production amplitude a non-zero phase on the radion pole. Markers indicate the benchmark masses.}
\label{fig:formfactor}
\end{figure*}

Using
\begin{equation}
A_{1/2}(\tau)=\frac{2}{\tau^2}\left[\tau+(\tau-1)f(\tau)\right],\qquad \tau=\frac{\shat}{4m_t^2},
\end{equation}
with
\begin{equation}
f(\tau)=
\begin{cases}
\arcsin^2\!\sqrt\tau, & \tau\le1,\\[4pt]
-\dfrac14\left[\ln\!\left(\dfrac{1+\sqrt{1-1/\tau}}{1-\sqrt{1-1/\tau}}\right)-i\pi\right]^2, & \tau>1,
\end{cases}
\end{equation}
the production coefficient used in the event-level mapping is
\begin{equation}
\boxed{C_\phi(\shat)=\frac{2m_t^2}{\Lr^2}F_\phi(\shat).}
\end{equation}
At the 800 GeV benchmark,
\begin{equation}
F_\phi(800^2)=7.226073779680290+0.705840939914568\,i,
\end{equation}
for $m_t=172.5$ GeV. Figure~\ref{fig:formfactor} shows the real and absorptive parts, magnitude, and phase over the full benchmark range.

\section{QCD production and radion--continuum interference}
\subsection{Partonic channels and coupling structure}
At the LHC, the Standard Model $t\bar t$ continuum receives contributions from both $gg$ and $q\bar q$ initial states. The radion signal considered here is produced through gluon fusion, so amplitude-level signal--background interference occurs only with the $gg\to t\bar t$ continuum; the $q\bar q$ channel contributes to the background yield but not to this interference term. At the amplitude level, the radion numerator contains one factor of the $gg\phi$ coupling and one factor of the $\phi t\bar t$ coupling and therefore scales as $\Lr^{-2}$. Consequently the interference numerator is linear in the radion amplitude, whereas the resonance-squared numerator is quadratic. The physical event-level mapping also rescales the width as $\Gamma_\phi\propto\Lr^{-2}$, so no fixed-power approximation is imposed on the final line shape near the pole.

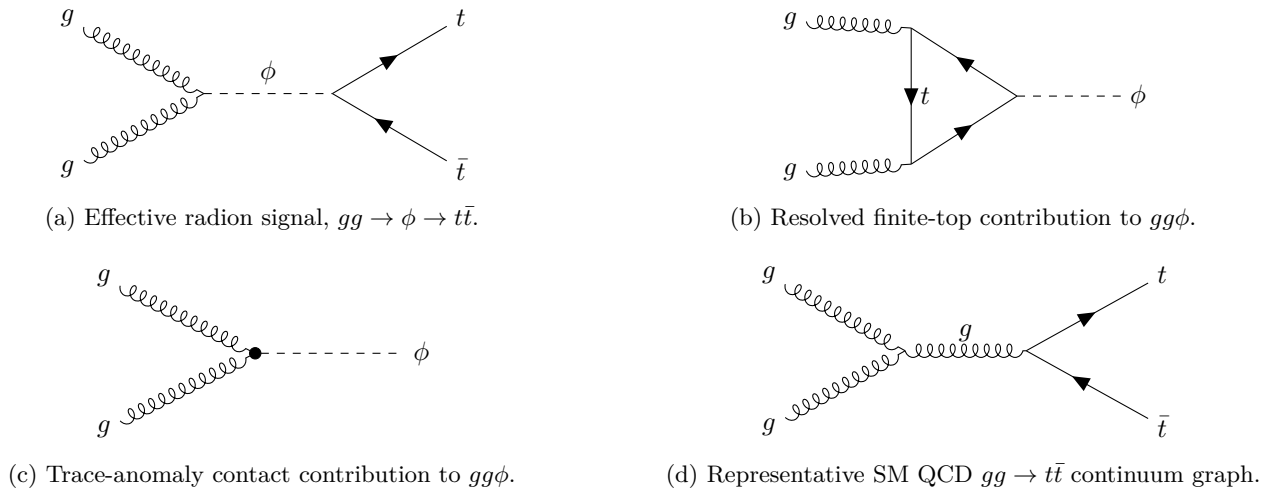
\begin{figure*}[tbp]
\centering
\begin{subfigure}{0.47\textwidth}
\centering
\begin{tikzpicture}
\begin{feynman}
\vertex (g1) at (0,1) {$g$};
\vertex (g2) at (0,-1) {$g$};
\vertex (a) at (1.8,0);
\vertex (b) at (3.5,0);
\vertex (t) at (5.2,1) {$t$};
\vertex (tb) at (5.2,-1) {$\bar t$};
\diagram*{(g1)--[gluon](a),(g2)--[gluon](a),(a)--[scalar,edge label={$\phi$}](b),(b)--[fermion](t),(b)--[anti fermion](tb)};
\end{feynman}
\end{tikzpicture}
\caption{Effective radion signal, $gg\to\phi\to t\bar t$.}
\end{subfigure}\hfill
\begin{subfigure}{0.47\textwidth}
\centering
\begin{tikzpicture}
\begin{feynman}
\vertex (g1) at (0,1) {$g$};
\vertex (g2) at (0,-1) {$g$};
\vertex (a) at (1.6,0.9);
\vertex (b) at (1.6,-0.9);
\vertex (c) at (3.0,0);
\vertex (p) at (4.6,0) {$\phi$};
\diagram*{(g1)--[gluon](a),(g2)--[gluon](b),(a)--[fermion,edge label={$t$}](b),(b)--[fermion](c),(c)--[fermion](a),(c)--[scalar](p)};
\end{feynman}
\end{tikzpicture}
\caption{Resolved finite-top contribution to $gg\phi$.}
\end{subfigure}

\vspace{0.8em}
\begin{subfigure}{0.47\textwidth}
\centering
\begin{tikzpicture}
\begin{feynman}
\vertex (g1) at (0,1) {$g$};
\vertex (g2) at (0,-1) {$g$};
\vertex[dot] (a) at (2,0) {};
\vertex (p) at (4.2,0) {$\phi$};
\diagram*{(g1)--[gluon](a),(g2)--[gluon](a),(a)--[scalar](p)};
\end{feynman}
\end{tikzpicture}
\caption{Trace-anomaly contact contribution to $gg\phi$.}
\end{subfigure}\hfill
\begin{subfigure}{0.47\textwidth}
\centering
\begin{tikzpicture}
\begin{feynman}
\vertex (g1) at (0,1) {$g$};
\vertex (g2) at (0,-1) {$g$};
\vertex (a) at (1.8,0);
\vertex (b) at (3.4,0);
\vertex (t) at (5.2,1) {$t$};
\vertex (tb) at (5.2,-1) {$\bar t$};
\diagram*{(g1)--[gluon](a),(g2)--[gluon](a),(a)--[gluon,edge label={$g$}](b),(b)--[fermion](t),(b)--[anti fermion](tb)};
\end{feynman}
\end{tikzpicture}
\caption{Representative SM QCD $gg\to t\bar t$ continuum graph.}
\end{subfigure}
\caption{Schematic amplitude content relevant for radion--QCD interference. The physical $gg\phi$ production amplitude is the coherent sum of the resolved top loop and the trace-anomaly contact term. Only the $gg$ component of the Standard Model continuum interferes with the gluon-fusion radion amplitude.}
\label{fig:feynman}
\end{figure*}

Figure~\ref{fig:feynman} summarizes the ingredients relevant to the coherent $gg$ channel. The effective signal diagram in panel (a) is resolved into the finite-top triangle and the trace-anomaly contact contribution in panels (b) and (c). Panel (d) shows one representative LO QCD continuum graph; the full $gg\to t\bar t$ background contains the usual $s$-, $t$-, and $u$-channel diagrams.

\subsection{Coherent decomposition}
The partonic amplitude is
\begin{equation}
\begin{aligned}
\mathcal M&=\mathcal M_{\rm QCD}+\mathcal M_\phi,\\
\mathcal M_\phi&=\mathcal M_{gg\phi}
\frac{\mathcal M_{\phi t\bar t}}
{\shat-\mph^2+i\mph\Gamma_\phi}.
\end{aligned}
\end{equation}
Squaring gives
\begin{equation}
|\mathcal M|^2=\underbrace{|\mathcal M_{\rm QCD}|^2}_{B}+\underbrace{2\operatorname{Re}(\mathcal M_{\rm QCD}^*\mathcal M_\phi)}_{I}+\underbrace{|\mathcal M_\phi|^2}_{S}.
\end{equation}

For the equal-helicity gluon configuration relevant to the scalar state, the analytic interference contract used for validation can be written
\begin{equation}
\frac{d\hat\sigma_I}{dz}=-\frac{\as^2}{64\pi}\frac{\beta^3}{1-\beta^2z^2}\,\operatorname{Re}[P(\shat)C_\phi(\shat)],
\end{equation}
with
\begin{equation}
P(\shat)=\frac{1}{\shat-\mph^2+i\mph\Gamma_\phi},\qquad \beta=\sqrt{1-\frac{4m_t^2}{\shat}}.
\end{equation}
After angular integration,
\begin{equation}
\boxed{\hat\sigma_I=-\frac{\as^2}{32\pi}\beta^2\operatorname{artanh}(\beta)\,\operatorname{Re}[P(\shat)C_\phi(\shat)].}
\end{equation}
At the exact pole, $P=-i/(\mph\Gamma_\phi)$. A purely real point-like coefficient gives $\operatorname{Re}(PC)=0$, whereas for $C=|C|e^{i\delta}$,
\begin{equation}
\operatorname{Re}(PC)=-\frac{|C|}{\mph\Gamma_\phi}\sin\delta,
\end{equation}
so the absorptive loop phase generates non-zero on-pole interference.

The peak--dip ordering can be checked analytically without relying on the event generator. Writing
\begin{equation}
 C_\phi=C_R+iC_I,\qquad \Delta_s=\shat-\mph^2,
\end{equation}
one obtains
\begin{equation}
\boxed{
\operatorname{Re}[P(\shat)C_\phi(\shat)]
=\frac{\Delta_s C_R+\mph\Gamma_\phi C_I}
{\Delta_s^2+\mph^2\Gamma_\phi^2}.}
\label{eq:peakdipsign}
\end{equation}
For the convention used here, $C_R>0$ and $C_I>0$ above the top threshold. The overall minus sign in the interference kernel therefore gives a positive dispersive contribution below the pole and a negative one above it. The absorptive term shifts the zero to
\begin{equation}
\Delta_s^{\rm zero}=-\mph\Gamma_\phi\frac{C_I}{C_R}.
\end{equation}
At $m_\phi=800$ GeV, $\Gamma_\phi=2.44$ GeV and $C_I/C_R=0.705841/7.226074$, corresponding to a mass shift $\Delta m\simeq\Delta_s/(2m_\phi)\simeq-0.12$ GeV. The zero is thus slightly below the pole and well below the 2 GeV truth-bin width. This provides a direct reader-facing check of the relative sign between the trace-anomaly and resolved-loop pieces.

\subsection{Common-event phase basis}
To reconstruct the coherent radion contribution from common generator events, introduce a scalar production phase $\varphi$ and define
\begin{equation}
w(\varphi)=|\mathcal M_B+e^{i\varphi}\mathcal M_S|^2.
\end{equation}
The five weights $w_0,w_{180},w_{90},w_{270},w_{\rm null}$ invert to
\begin{align}
B&=w_{\rm null},\\
S&=\frac{w_0+w_{180}}{2}-B,\\
I_0&=\frac{w_0-w_{180}}{2},\\
I_{90}&=\frac{w_{90}-w_{270}}{2}.
\end{align}
Here $I_0$ and $I_{90}$ are the in-phase and quadrature interference components. Over-constrained checks such as
\begin{equation}
I_{45}=\frac{I_0+I_{90}}{\sqrt2},\qquad I_{34}=0.6I_0+0.8I_{90}
\end{equation}
close numerically on common events.

The physical radion is then reconstructed event by event through
\begin{equation}
R(\shat;\Lr)=\frac{C_\phi(\shat;\Lr)}{C_{\rm ref}}\,
\frac{\shat-\mph^2+i\mph\Gamma_0}{\shat-\mph^2+i\mph\Gamma_\phi(\Lr)},
\end{equation}
where $C_{\rm ref}$ is the normalization of the generic scalar amplitude used to construct the phase basis. In the frozen \texttt{htt4} convention it is
\begin{equation}
 C_{\rm ref}=-32\pi^2m_tY_{Ht}\left(\frac{C_{\phi G}}{\Lambda}\right)c_c.
\end{equation}
With $m_t=172.5$ GeV, $Y_{Ht}=0.7$, $C_{\phi G}=-0.00938$, $\Lambda=1000$ GeV, and $c_c=-3$, this gives
\begin{equation}
 \boxed{C_{\rm ref}=-1.073151300559473.}
\end{equation}
It is therefore a basis-normalization constant, not a physical radion coupling; the physical prediction enters through the ratio $C_\phi/C_{\rm ref}$ together with the propagator ratio. The reconstructed components are
\begin{equation}
\boxed{S_{\rm rad}=|R|^2S,\qquad I_{\rm rad}=\operatorname{Re}(R)I_0+\operatorname{Im}(R)I_{90}.}
\end{equation}

\section{Generator implementation and validation}
The phase-basis samples are generated with \textsc{MadGraph5\_aMC@NLO} v3.5.5~\cite{MG5}, using the \texttt{htt4} generic CP-even scalar model as a computational amplitude basis and the process command \texttt{generate p p > t t\string~ QCD=2 QED=1 EFT=1}. The reference scalar is not interpreted directly as a radion: the physical anomaly-plus-loop coefficient and running width are imposed through the event-level mapping above. Table~\ref{tab:generator} records the frozen run-card provenance.

\begin{table*}[tbp]
\centering
\caption{Generator provenance of the 14 TeV phase-basis samples used in this revision.}
\label{tab:generator}
\begin{tabular}{p{0.28\textwidth}p{0.64\textwidth}}
\toprule
item & frozen setting \\
\midrule
beam energy & $E_{\rm beam}=7.0$ TeV per proton, i.e. $\sqrt{s}=14$ TeV \\
PDF & MadGraph internal \texttt{nn23lo1}, corresponding to \texttt{NNPDF23\_lo\_as\_0130\_qed} member 0 with $\alpha_s(M_Z)=0.130$; the run-card \texttt{lhaid=230000} is inactive because \texttt{pdlabel} is not \texttt{lhapdf} \\
scales & non-fixed renormalization and factorization scales; MadGraph default dynamical prescription (\texttt{dynamical\_scale\_choice=-1}) with \texttt{scalefact=1} \\
parton-level cuts & no additional $\sqrt{\hat s}$, $p_T$, or $\eta$ cuts in the stored run card \\
800 GeV sample & 500,000 unweighted events, random seed 27 \\
\bottomrule
\end{tabular}
\end{table*}

For the 800 GeV sample, generator systematic reweights were produced with $\mu_R$ and $\mu_F$ factors $0.5,1,2$ and an NNPDF error-set scan. The reweighting metadata use \texttt{NNPDF23\_lo\_as\_0130\_qed} LHAPDF ID 247000 for the central member and 247001--247100 for the replicas. The inclusive generator-level audit gives a scale envelope of $+30.3\%/-21.8\%$ and a PDF variation of $\pm4.38\%$. If that inclusive scale envelope were treated as an independent normalization uncertainty on an interference-dominated signal with $s\propto\Lr^{-2}$, it would correspond schematically to about $+14.1\%/-11.6\%$ in the reach. That is intentionally not used as a final uncertainty: common scale and PDF variations affect both the QCD continuum and the coherent deformation and cancel substantially in a ratio treatment. Indeed, the earlier common-event reweighting study, where $B,S,I_0,I_{90}$ were varied coherently, gave only a $+2.68\%/-2.86\%$ legacy-Gaussian reach envelope and a 1.13\% PDF-replica RMS. These diagnostics motivate the selected-yield deformation formulation introduced below; neither result is promoted to a complete NLO theory uncertainty.

An independent native validation uses MadGraph's loop-induced machinery~\cite{LoopInduced} and a UFO model prepared with FeynRules~\cite{FeynRules}, in which the anomaly contact contribution is assigned to the same split-order sector as the loop amplitude. The extracted native complex amplitudes at 750, 800, and 850 GeV agree with the analytic contact-plus-loop construction at relative precision between $10^{-16}$ and $10^{-15}$. The validated complex loop anchors are shown in Table~\ref{tab:loopanchors}.

\begin{table}[htbp]
\centering
\caption{Three-mass form-factor anchors used in the native amplitude validation.}
\label{tab:loopanchors}
\resizebox{\columnwidth}{!}{%
\begin{tabular}{lcc}
\toprule
$\mph$ [GeV] & $A_{1/2}$ & $F_\phi=7+\tfrac12A_{1/2}$\\
\midrule
750 & $0.581232+1.480056i$ & $7.290616+0.740028i$\\
800 & $0.452148+1.411682i$ & $7.226074+0.705841i$\\
850 & $0.346618+1.340821i$ & $7.173309+0.670411i$\\
\bottomrule
\end{tabular}}
\end{table}

The six event samples contain 100k events at 400, 600, 1000, 1200, and 1500 GeV and 500k events at 800 GeV. The 800 GeV sample is used for high-statistics interference, resolution, scale, PDF, and component studies.

\section{Truth morphology and detector response}
\subsection{Truth-level peak--dip structure}
\begin{figure*}[tbp]
\centering
\begin{subfigure}{0.49\textwidth}
\includegraphics[width=\textwidth]{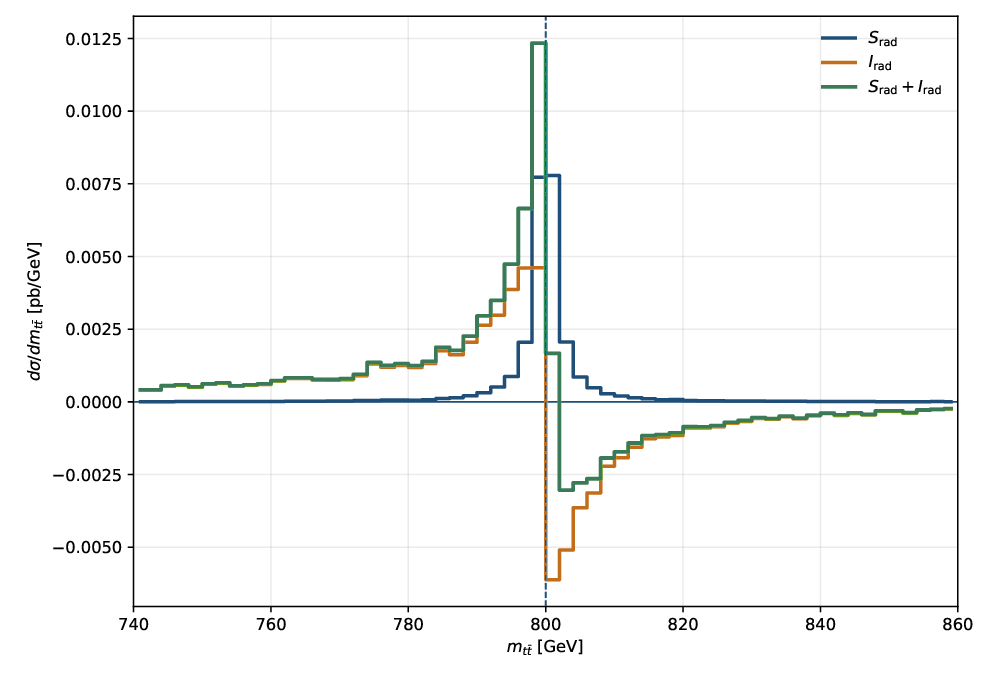}
\caption{Background, signal-squared, interference, and coherent components.}
\end{subfigure}\hfill
\begin{subfigure}{0.49\textwidth}
\includegraphics[width=\textwidth]{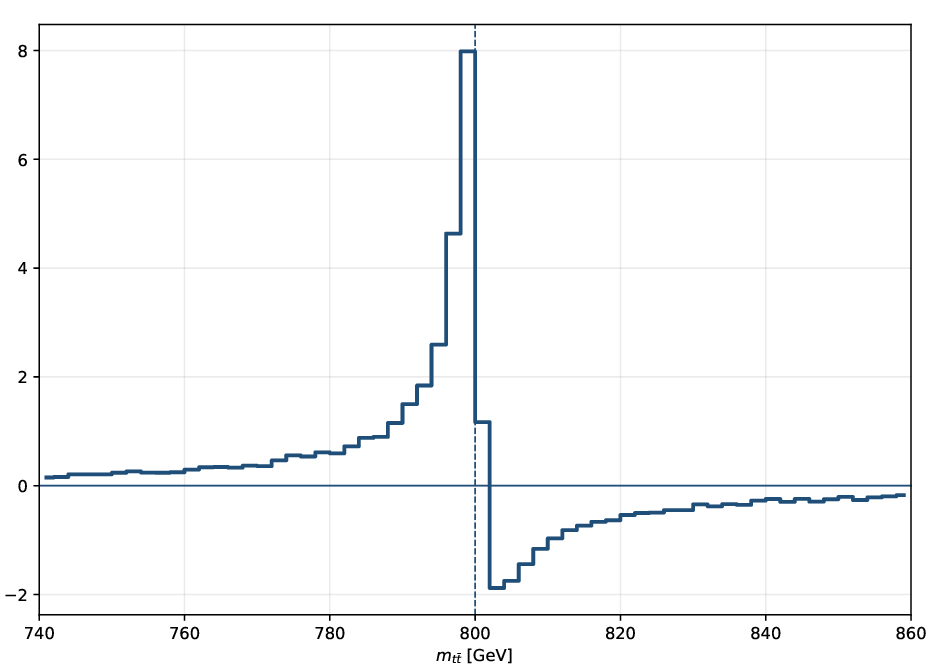}
\caption{Fractional coherent deformation.}
\end{subfigure}
\caption{Validated truth-level 800 GeV morphology. The radion contribution is sign-changing and interference dominated rather than a conventional positive bump.}
\label{fig:truth800}
\end{figure*}

At $\mph=800$ GeV and $\Lr=3$ TeV the validated 500k sample exhibits a sharp asymmetric peak--dip in $(S+I)/B$. With the 2 GeV truth bins used in Fig.~\ref{fig:truth800}, the positive maximum is approximately $+7.98\%$ at 799 GeV and the negative minimum is approximately $-1.88\%$ at 803 GeV. Because the bin width is comparable to the intrinsic 2.44 GeV width, these local extrema are explicitly binning dependent; the integrated interference and the detector-smeared morphology are the more stable observables. The positive lobe is larger because the resonance-squared term partially fills the destructive side of the interference.

\subsection{ATLAS-anchored phenomenological response}
The detector model uses two independent public ATLAS inputs. For the reconstructed-$\mtt$ response we use the heavy-neutral-Higgs $t\bar t$ analysis~\cite{ATLAS2024}, which reports a resolved-topology mass resolution of about 12\% at $\mtt=400$ GeV improving to about 10\% at 1 TeV, and a merged-topology resolution of about 10\%. These measurements motivate the mass-dependent Gaussian response used below.

\begin{figure*}[tbp]
\centering
\includegraphics[width=0.90\textwidth]{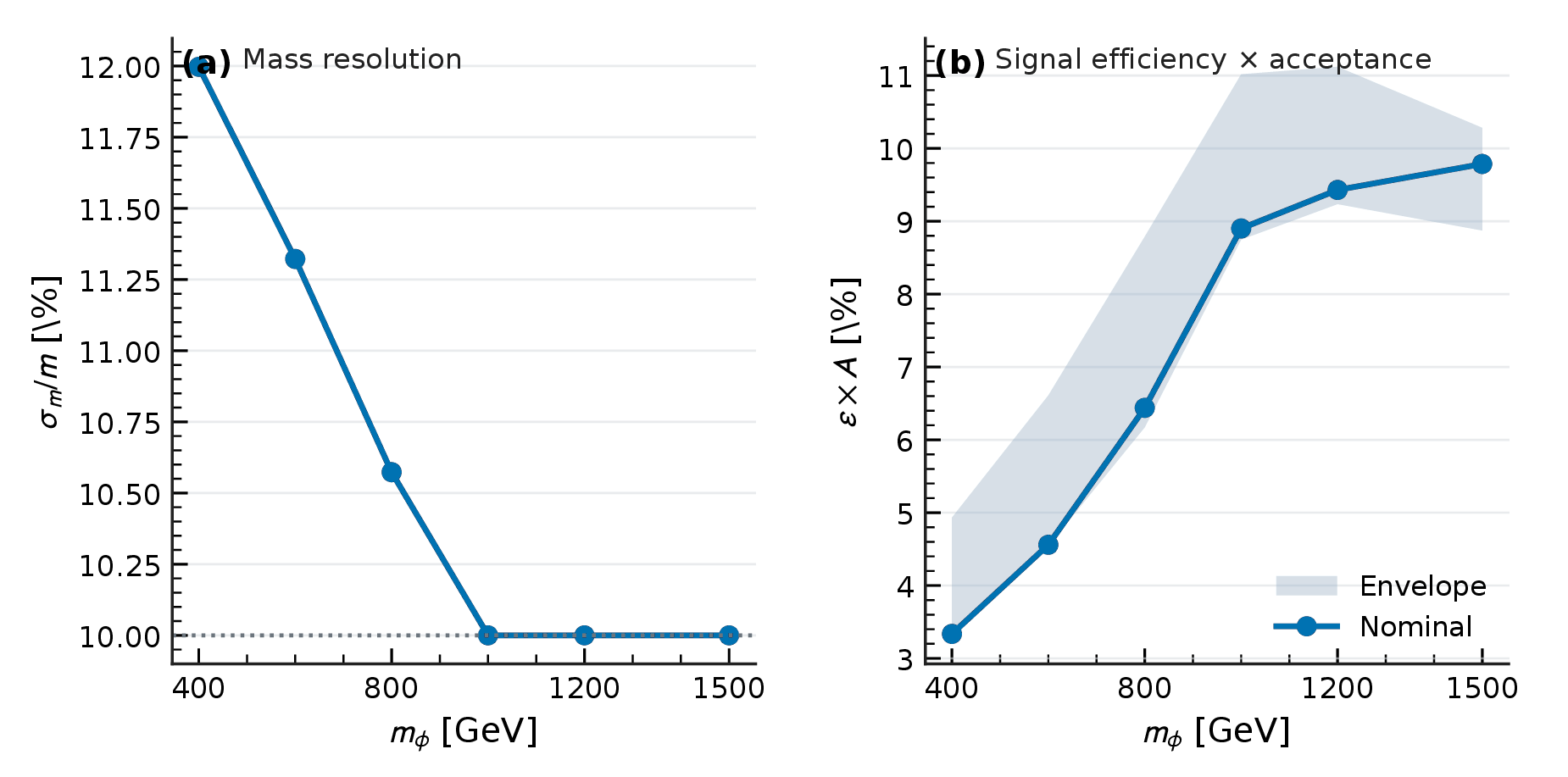}
\caption{Detector inputs used in the projection. The reconstructed-$\mtt$ resolution is anchored to the ATLAS heavy-neutral-Higgs search~\cite{ATLAS2024}. The efficiency-times-acceptance points are derived from the inclusive resolved+merged curves of the ATLAS resonance study~\cite{ATLASRes2025,ATLASResHEPData}; the band spans the three public benchmark signal models and is used only as a signal-model envelope.}
\label{fig:detinputs}
\end{figure*}

The signal efficiency-times-acceptance is derived instead from the public ATLAS $t\bar t$-resonance study~\cite{ATLASRes2025} and its HEPData record~\cite{ATLASResHEPData}. We use the inclusive resolved+merged efficiency curves in Fig.~2 of that study for the three benchmark signal models $Z'_{\rm TC2}$, $G_{\rm KK}$, and $g_{\rm KK}$. At each radion mass, the three curves are linearly interpolated; their median defines the nominal $\epsilon A$, while their minimum and maximum define the low/high signal-model envelope. This construction is an experimentally motivated phenomenological proxy for topology and signal-model dependence. It is not an ATLAS radion efficiency and the envelope is not interpreted as an official experimental systematic uncertainty. The resulting inputs are summarized in Table~\ref{tab:detector} and shown in Fig.~\ref{fig:detinputs}.

\begin{table*}[tbp]
\centering
\caption{Nominal response inputs used in the projection.}
\label{tab:detector}
\begin{tabular}{rcccc}
\toprule
$\mph$ [GeV] & $\sigma_m/m$ & nominal $\epsilon A$ & $\epsilon A$ envelope & likelihood window [GeV] \\
\midrule
400 & 11.997\% & 3.336\% & 3.316--4.933\% & 350--500 \\
600 & 11.322\% & 4.559\% & 4.545--6.610\% & 450--750 \\
800 & 10.573\% & 6.439\% & 6.172--8.789\% & 600--1000 \\
1000 & 10.0\% & 8.901\% & 8.749--11.017\% & 750--1250 \\
1200 & 10.0\% & 9.430\% & 9.236--11.120\% & 900--1500 \\
1500 & 10.0\% & 9.786\% & 8.873--10.283\% & 1125--1875 \\
\bottomrule
\end{tabular}
\end{table*}

\begin{figure*}[tbp]
\centering
\includegraphics[width=0.98\textwidth]{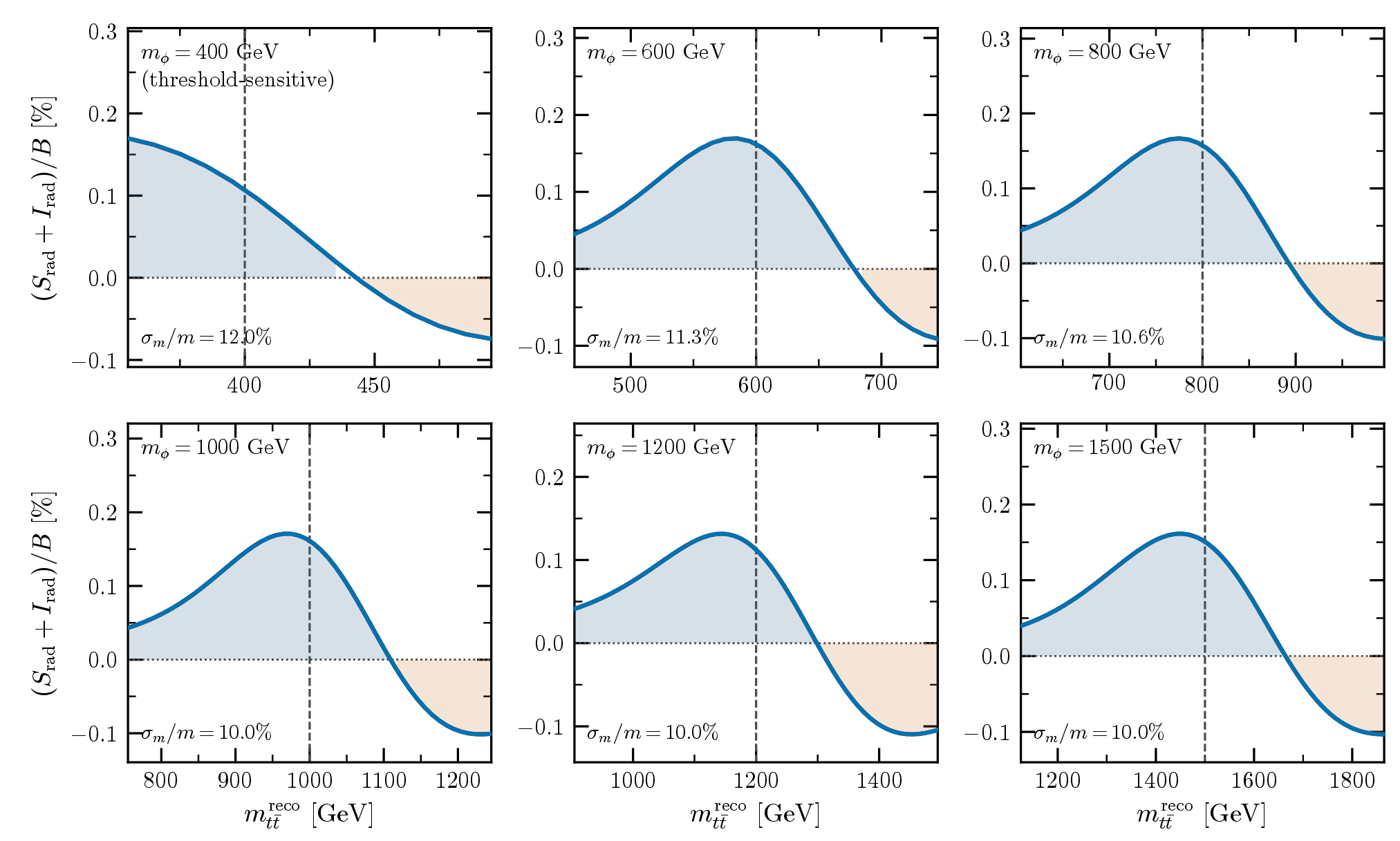}
\caption{Detector-smeared radion--QCD deformation $(S_{\rm rad}+I_{\rm rad})/B$ at $\Lr=3$ TeV for the six benchmark masses, using the nominal mass-dependent response. The 400 GeV panel is explicitly threshold sensitive.}
\label{fig:multimass}
\end{figure*}

A Gaussian response matrix maps truth-bin center $m_i$ into reconstructed bin $j$,
\begin{equation}
\begin{aligned}
R_{ij}=\frac12\Bigg[&\operatorname{erf}\!\left(\frac{m_{j+1}-m_i}{\sqrt2\sigma_i}\right)
-\operatorname{erf}\!\left(\frac{m_j-m_i}{\sqrt2\sigma_i}\right)\Bigg],\\
&\sigma_i=(\sigma_m/m)m_i.
\end{aligned}
\end{equation}
After smearing, the sharp truth-level structure becomes a broad sub-percent deformation. Figure~\ref{fig:multimass} shows the reconstructed line shapes at all six masses.

\section{Statistical inference}
\subsection{Expected yields and ATLAS background anchor}
The measured background and the simulated radion deformation enter the projection through two complementary objects. First, the likelihood used for the quoted reach uses the absolute reconstructed signed radion template,
\begin{equation}
s_i^{\rm abs}=H^{\rm reco}_{S+I,i}(\Lr)\,\mathcal L\,(10^3\ \mathrm{fb/pb})\,(\epsilon A),
\label{eq:signalabs}
\end{equation}
with $\mathcal L=3000$ fb$^{-1}$. The vector is signed because interference can reduce the expectation in individual bins. The 14 TeV phase-basis samples are used directly, so no additional collision-energy rescaling is applied to $H^{\rm reco}_{S+I}$.

For the background, the MadGraph continuum retains the fine reconstructed-$\mtt$ shape, while the normalization in each likelihood window $W$ is replaced by the public ATLAS post-fit selected-background yield obtained by summing the exclusive resolved and merged signal regions in the resonance analysis~\cite{ATLASRes2025,ATLASResHEPData}. Defining
\begin{equation}
\widehat H^{\rm reco}_{B,i}=\frac{H^{\rm reco}_{B,i}}{\sum_{j\in W}H^{\rm reco}_{B,j}},
\end{equation}
the selected-background transfer is
\begin{equation}
\begin{aligned}
N_B^{\rm ATLAS\to HL}(W)&=N_B^{\rm ATLAS,13}(W)\\
&\times\left(\frac{3000}{140}\right)
\left(\frac{\sigma_{t\bar t}^{14}}{\sigma_{t\bar t}^{13}}\right).
\end{aligned}
\label{eq:bkgtransfer}
\end{equation}
For $m_t=172.5$ GeV, the LHC Top Working Group Top++ references give $\sigma_{t\bar t}^{13}=833.9$ pb and $\sigma_{t\bar t}^{14}=985.7$ pb~\cite{LHCTopWG,CzakonFiedlerMitov2013}. Hence
\begin{equation}
 \frac{\sigma_{t\bar t}^{14}}{\sigma_{t\bar t}^{13}}=1.18204,
 \qquad
 \left(\frac{3000}{140}\right)\left(\frac{985.7}{833.9}\right)=25.3293.
\end{equation}
The expected background in this implementation is therefore
\begin{equation}
\boxed{b_i=N_B^{\rm ATLAS\to HL}\,\widehat H^{\rm reco}_{B,i}.}
\label{eq:bkgtemplate}
\end{equation}
The numerical transfer used at each mass point is given in Table~\ref{tab:bkgtransfer}. The inclusive 13-to-14 TeV factor is applied uniformly within each local window; a window-differential transfer would be preferable at high mass.

\begin{table*}[tbp]
\centering
\caption{Numerical selected-background normalization used in the likelihood. $N_B^{\rm ATLAS,13}$ is the summed post-fit background in the exclusive ATLAS resolved and merged signal regions; the final column applies the common factor 25.3293 from Eq.~\eqref{eq:bkgtransfer}.}
\label{tab:bkgtransfer}
\begin{tabular}{rccc}
\toprule
$\mph$ [GeV] & fit window [GeV] & $N_B^{\rm ATLAS,13}$ & $N_B^{\rm ATLAS\to HL}$ \\
\midrule
400  & 350--500   & 1,896,682 & 48,041,719 \\
600  & 450--750   & 1,601,330 & 40,560,653 \\
800  & 600--1000  & 679,037   & 17,199,561 \\
1000 & 750--1250  & 308,652   & 7,817,949 \\
1200 & 900--1500  & 137,781   & 3,489,897 \\
1500 & 1125--1875 & 43,720    & 1,107,405 \\
\bottomrule
\end{tabular}
\end{table*}

\subsubsection*{Selected-yield deformation formulation}
The asymmetry between an absolute simulated signal and a data-anchored background can be reduced by normalizing the BSM contribution as a fractional deformation. Define
\begin{equation}
 r_i(\Lr)=\frac{H^{\rm reco}_{S+I,i}(\Lr)}{H^{\rm reco}_{B,i}}.
\label{eq:deformationratio}
\end{equation}
If $b_i^{t\bar t}$ denotes the selected $t\bar t$ component of the measured background, the corresponding selected-yield deformation is
\begin{equation}
\boxed{s_i^{\rm ratio}=r_i(\Lr)\,b_i^{t\bar t},\qquad
\mu_i=b_i^{\rm non-tt}+b_i^{t\bar t}+s_i^{\rm ratio}.}
\label{eq:rationorm}
\end{equation}
This formulation has three advantages. Common powers of $\alpha_s$, a large part of the PDF dependence, and the dominant inclusive QCD $K$ factor cancel between numerator and denominator. To the extent that selection depends mainly on reconstructed $\mtt$, the continuum selection efficiency is already contained in $b_i^{t\bar t}$, leaving only a residual acceptance correction associated with the different scalar angular structure. Finally, the ratio is much less sensitive to the 13-to-14 TeV luminosity change than either absolute cross section separately.

Equation~\eqref{eq:rationorm} should not, however, be implemented by multiplying the deformation by the total selected background without qualification: non-$t\bar t$ backgrounds do not carry the radion deformation, and the scalar and continuum angular distributions need not have identical acceptance. The public inputs used here have not been propagated into a category-resolved $b_i^{t\bar t}$ template with an event-level angular acceptance correction. For this reason the fully reprofiled reach numbers quoted below remain the absolute-normalization results of Eq.~\eqref{eq:signalabs}.

To quantify the normalization dependence with the available public inputs, we additionally run a deliberately simplified \emph{selected-background deformation diagnostic},
\begin{equation}
\boxed{s_i^{\rm sel}=r_i(\Lr)\,b_i,}
\label{eq:selbkgdiag}
\end{equation}
where $b_i$ is the total ATLAS-anchored selected-background template of Eq.~\eqref{eq:bkgtemplate}, rather than the $t\bar t$-only component required by Eq.~\eqref{eq:rationorm}. Because the present hybrid $b_i$ uses the same reconstructed MadGraph continuum shape $H^{\rm reco}_{B,i}$ as the denominator of $r_i$, Eq.~\eqref{eq:selbkgdiag} leaves the signed peak--dip morphology unchanged and isolates the mass-dependent normalization-anchor effect. In the implementation this is algebraically equivalent to multiplying the absolute signal template by the corresponding continuum anchor factor $r_B(m_\phi)$ while keeping the final selected background fixed. The exact Poisson profiler is rerun at all six masses with the same $\fshape=0.1\%$, $\ell=50$ GeV nuisance model. The results are presented only as a robustness diagnostic and are not identified with the category-resolved likelihood of Eq.~\eqref{eq:rationorm}.

\subsection{Correlated background-shape model}
The correlated shape covariance is
\begin{equation}
\boxed{K^{\rm shape}_{ij}=\fshape^2 b_i b_j\exp\!\left[-\frac{(m_i-m_j)^2}{2\ell^2}\right].}
\end{equation}
The benchmark scenario uses
\begin{equation}
\boxed{\fshape=0.1\%,\qquad \ell=50\ \mathrm{GeV}.}
\end{equation}
This choice should be interpreted as a sensitivity benchmark for residual background-shape control rather than as an experimentally measured HL-LHC uncertainty. Factorizing $K^{\rm shape}=LL^T$, the profiled expectation is
\begin{equation}
\boxed{\mu_i(\alpha,z)=\alpha b_i+s_i+(Lz)_i,}
\end{equation}
where $\alpha>0$ is a free overall background normalization and $z$ is a vector of standard-normal nuisance parameters. The normalization freedom is therefore handled by $\alpha$ and is not duplicated inside $K^{\rm shape}$.

For the background-only Asimov data set $n_i=b_i$, the profiled objective is
\begin{equation}
\boxed{q_A(\alpha,z)=2\sum_i\left[\mu_i-b_i+b_i\ln\frac{b_i}{\mu_i}\right]+z^Tz.}
\end{equation}
For fixed normalization, $\alpha=1$; otherwise $\alpha$ and $z$ are profiled simultaneously.

\subsection{Exact-Hessian profiling and asymptotic \texorpdfstring{$CL_s$}{CLs}}
Defining
\begin{equation}
g_{\mu,i}=2\left(1-\frac{b_i}{\mu_i}\right),\qquad W_i=\frac{2b_i}{\mu_i^2},
\end{equation}
the gradient is
\begin{equation}
g_\alpha=b^Tg_\mu,\qquad g_z=L^Tg_\mu+2z,
\end{equation}
and the exact Hessian blocks are
\begin{equation}
\begin{aligned}
H_{\alpha\alpha}&=\sum_iW_ib_i^2,\\
H_{\alpha z}&=(Wb)^TL,\qquad
H_{zz}=L^TWL+2I.
\end{aligned}
\end{equation}
A Newton step with backtracking enforces $\alpha>0$, $\mu_i>0$, and sufficient decrease. Convergence is checked using the final Newton correction.

\begin{figure}[htbp]
\centering
\includegraphics[width=\columnwidth]{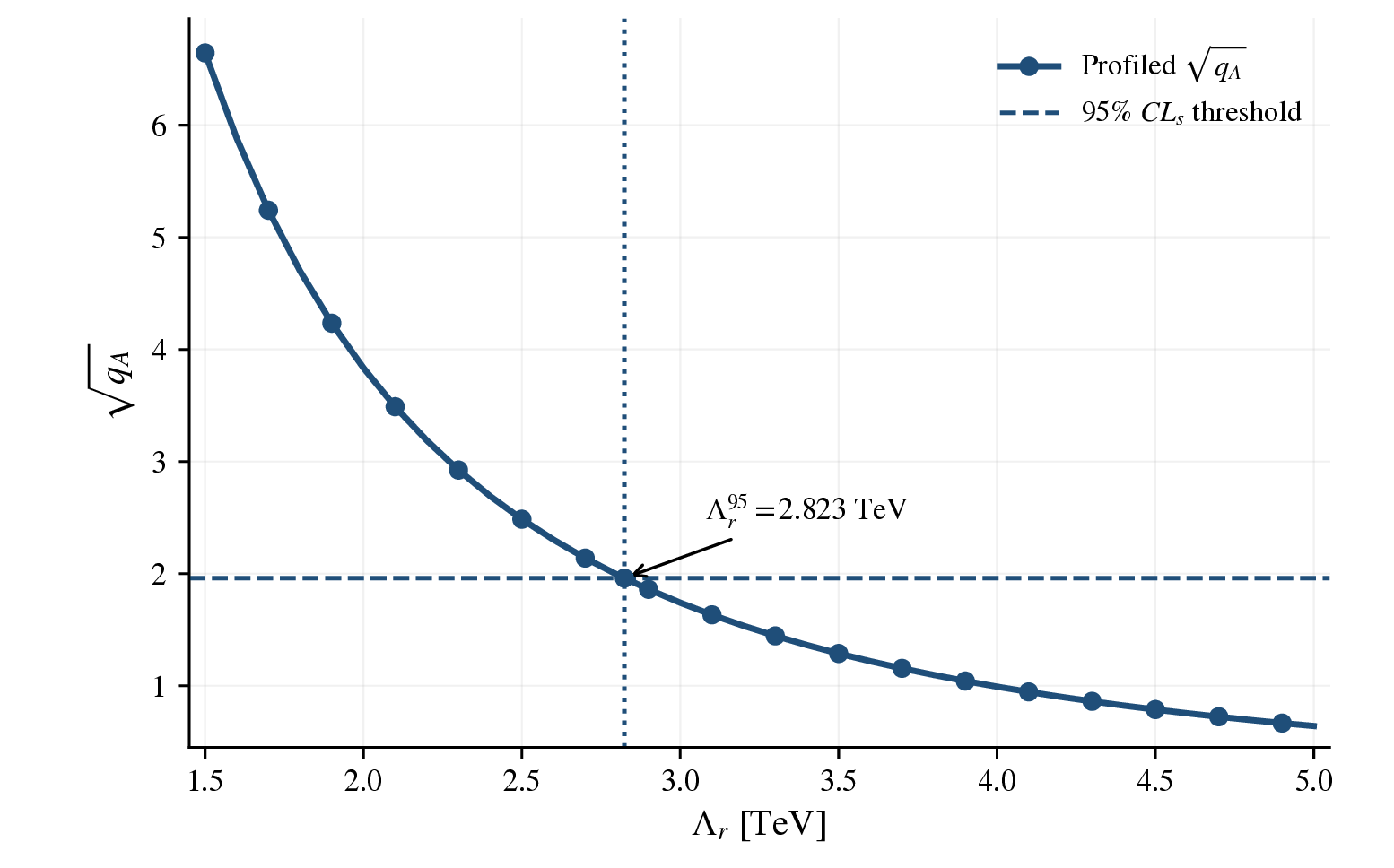}
\caption{Profiled 800 GeV Asimov $\sqrt{q_A}$ scan for the ATLAS-anchored background model. The median expected 95\% $\CLs$ threshold is crossed at $\Lr^{95}=2.82$ TeV; the unrounded numerical root is retained only in the validation record.}
\label{fig:clsscan}
\end{figure}

For the background-only Asimov data set, the median asymptotic mapping is~\cite{Cowan2011}
\begin{equation}
\CLs=2\left[1-\Phi\left(\sqrt{q_A}\right)\right]=\operatorname{erfc}\left(\frac{\sqrt{q_A}}{\sqrt2}\right),
\end{equation}
so the expected 95\% $\CLs$ threshold is
\begin{equation}
\boxed{\sqrt{q_A}=1.959963984540054.}
\end{equation}
The reach is found by bracketing the monotonic crossing in $\Lr$ and solving with Brent's method. At 800 GeV and $\Lr=3$ TeV, the exact Poisson profiler reproduces the earlier Gaussian-covariance diagnostic to better than $10^{-2}\%$. The corresponding profiled $\sqrt{q_A}$ scan is shown in Fig.~\ref{fig:clsscan}.

\section{Projected HL-LHC sensitivity}
\begin{figure}[htbp]
\centering
\includegraphics[width=\columnwidth]{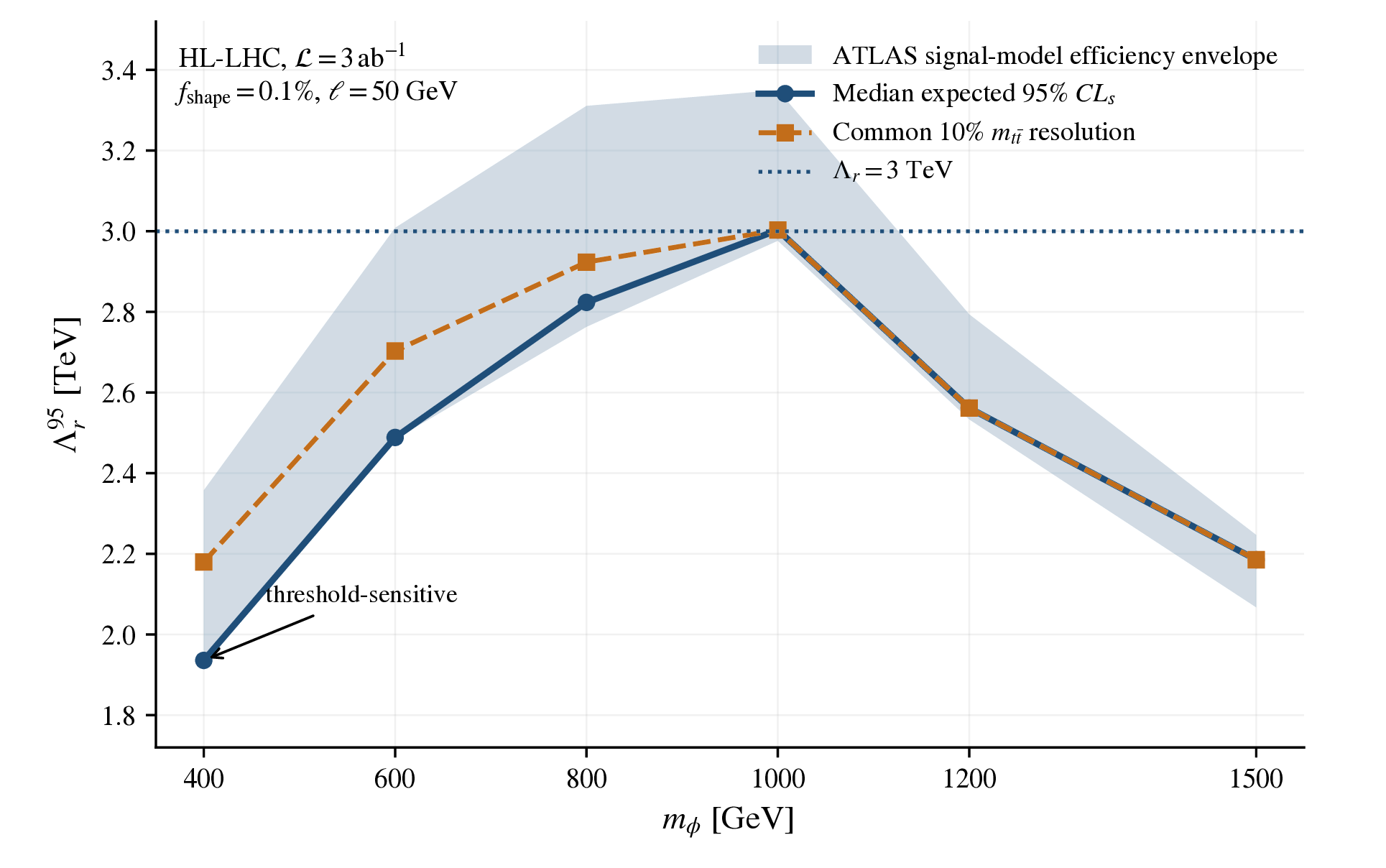}
\caption{Median expected 95\% $\CLs$ reach in $\Lr$ at the HL-LHC. The solid curve is the nominal profiled result. The shaded region is the signal-efficiency envelope, while the dashed curve imposes a common 10\% reconstructed-$\mtt$ resolution as a robustness variation. The 400 GeV point is shown as threshold sensitive.}
\label{fig:money}
\end{figure}

The final median expected 95\% $\CLs$ reaches are listed in Table~\ref{tab:reach} and shown in Fig.~\ref{fig:money}.

\begin{table*}[tbp]
\centering
\caption{Median expected asymptotic 95\% $\CLs$ reach in $\Lr$ with the ATLAS-anchored selected-background normalization. The 400 GeV point is threshold sensitive and is retained as an auxiliary benchmark; the primary interpretation is $\mph\ge600$ GeV.}
\label{tab:reach}
\begin{tabular}{rcc}
\toprule
$\mph$ [GeV] & fixed norm. [TeV] & profiled norm.+shape [TeV] \\
\midrule
400 & 2.02 & \textbf{1.94} \\
600 & 2.67 & \textbf{2.49} \\
800 & 3.11 & \textbf{2.82} \\
1000 & 3.45 & \textbf{3.00} \\
1200 & 2.95 & \textbf{2.56} \\
1500 & 2.86 & \textbf{2.19} \\
\bottomrule
\end{tabular}
\end{table*}

The expected reach rises from 1.94 TeV at the threshold-sensitive 400 GeV benchmark to 2.49 TeV at 600 GeV and reaches approximately 3.00 TeV near $\mph=1$ TeV. It then decreases to 2.56 TeV at 1.2 TeV and 2.19 TeV at 1.5 TeV. The origin of this maximum can be made quantitative rather than attributed generically to several competing effects.

\subsection{Origin of the mass-dependent maximum near 1 TeV}
\label{sec:masspeak}
\begin{figure}[htbp]
\centering
\includegraphics[width=\columnwidth]{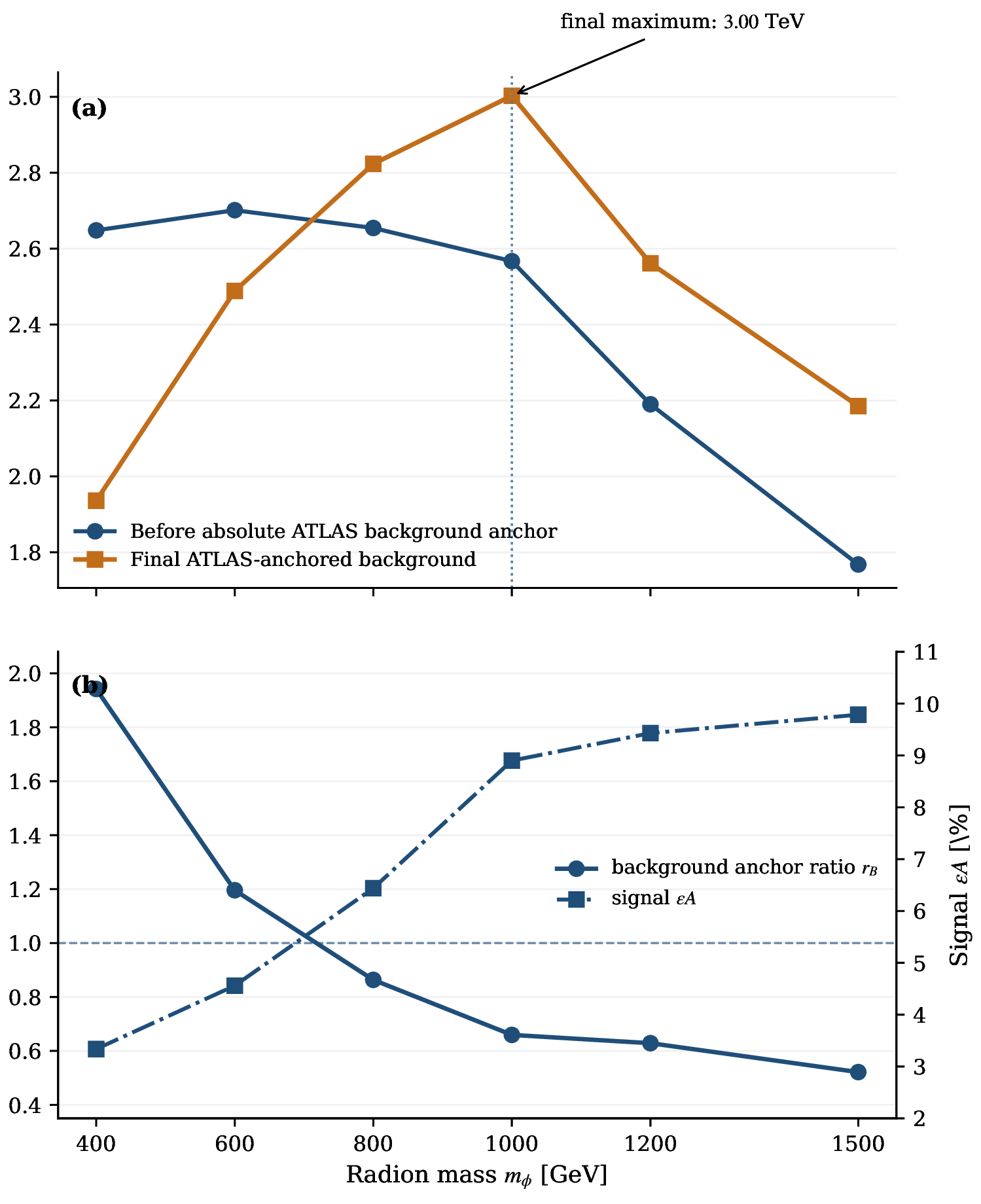}
\caption{Origin of the maximum near $m_\phi=1$ TeV. Top: the pre-anchor and final exact-Poisson profiled reaches. Bottom: the mass-dependent background normalization ratio $r_B$ and the signal efficiency-times-acceptance. The absolute ATLAS background anchor penalizes the low-mass windows ($r_B>1$) and reduces the continuum normalization for $m_\phi\ge800$ GeV ($r_B<1$), while the signal efficiency continues to rise rapidly up to about 1 TeV and then saturates.}
\label{fig:masspeakorigin}
\end{figure}

The interpretation in this subsection applies specifically to the absolute-normalization headline likelihood; the selected-background deformation diagnostic in Sec.~\ref{sec:selbkgdiag} shows that the precise location and height of the maximum are normalization-prescription dependent. The change in the mass dependence is primarily associated with replacing the earlier common-efficiency continuum normalization by the absolute selected-background anchor of Eq.~\eqref{eq:bkgtransfer}. To isolate this effect, Table~\ref{tab:masspeakdiag} compares the exact-Poisson profiled reach before and after that normalization update. The relevant quantities are compared in Fig.~\ref{fig:masspeakorigin}. The signal templates, mass-dependent response, efficiencies, and likelihood windows are otherwise the same analysis ingredients. The quantity $r_B$ is the multiplicative correction that maps the earlier continuum normalization in each window to the final ATLAS-selected yield.

\begin{table*}[tbp]
\centering
\caption{Diagnostic of the mass dependence of the reach. $N_B^{\rm HL}$ is the final selected-background target in the local likelihood window, and $r_B$ is the final-to-pre-anchor continuum normalization ratio. The ``pre-anchor'' column is retained only to diagnose the origin of the changed mass dependence; the final column is the result used elsewhere in the paper.}
\label{tab:masspeakdiag}
\small
\begin{tabular}{rccccc}
\toprule
$\mph$ [GeV] & $\epsilon A$ [\%] & $N_B^{\rm HL}$ [$10^6$] & $r_B$ & pre-anchor [TeV] & final [TeV] \\
\midrule
400  & 3.34 & 48.04 & 1.94 & 2.65 & 1.94 \\
600  & 4.56 & 40.56 & 1.20 & 2.70 & 2.49 \\
800  & 6.44 & 17.20 & 0.86 & 2.65 & 2.82 \\
1000 & 8.90 &  7.82 & 0.66 & 2.57 & 3.00 \\
1200 & 9.43 &  3.49 & 0.63 & 2.19 & 2.56 \\
1500 & 9.79 &  1.11 & 0.52 & 1.77 & 2.19 \\
\bottomrule
\end{tabular}
\end{table*}

The comparison shows that the 1 TeV maximum is not generated by an anomalous feature of the radion amplitude or by the normalization profiler. Before the absolute background anchor, the profiled reach already decreased gradually from 2.70 TeV at 600 GeV to 2.57 TeV at 1 TeV and 1.77 TeV at 1.5 TeV. The ATLAS anchor changes that balance in a strongly mass-dependent way: relative to the pre-anchor result, the final reach changes by $-7.9\%$ at 600 GeV, $+6.4\%$ at 800 GeV, $+17.0\%$ at 1 TeV, $+17.0\%$ at 1.2 TeV, and $+23.6\%$ at 1.5 TeV. The threshold-sensitive 400 GeV point is reduced by 26.9\%. These shifts track the crossing of $r_B$ through unity between 600 and 800 GeV.

Between 600 GeV and 1 TeV, two effects therefore act in the same favorable direction: the selected background in the local window falls from $40.6$ million to $7.82$ million events, while the signal efficiency rises from 4.56\% to 8.90\%. In the statistics-dominated limit one expects schematically $Z\sim s/\sqrt{b}$, so lowering a uniformly rescaled background by $r_B$ gives an approximate $r_B^{-1/2}$ gain; in the shape-systematics-dominated limit the covariance term scales as $K_{ij}\propto b_i b_j$, and the dependence can be stronger. The quoted reaches are obtained from the full profiled Poisson calculation rather than from either scaling approximation.

Above 1 TeV the efficiency gain largely saturates: $\epsilon A$ increases only from 8.90\% at 1 TeV to 9.79\% at 1.5 TeV. At the same time, the pre-anchor calculation already shows the loss of signal/interference leverage with increasing mass, reflecting the falling gluon-fusion parton luminosity together with the mass-dependent coherent line shape and its reconstruction. The further reduction of the selected background is no longer sufficient to compensate for that loss, and the final reach therefore falls from 3.00 TeV at 1 TeV to 2.19 TeV at 1.5 TeV. The maximum near 1 TeV is thus a crossover between the falling selected-background burden and rising efficiency at lower masses, and declining high-mass signal/interference leverage once the efficiency has largely saturated. The fact that the fixed-normalization reach also peaks at 1 TeV (3.45 TeV) confirms that normalization profiling modulates this pattern but does not create it. This interpretation is specific to the absolute-signal normalization of Eq.~\eqref{eq:signalabs}; the normalization diagnostic below tests how much of the detailed mass dependence survives when the BSM contribution is instead tied to the selected-background normalization.

\subsection{Selected-background deformation diagnostic}
\label{sec:selbkgdiag}
Using Eq.~\eqref{eq:selbkgdiag}, the exact Poisson likelihood is reprofiled at all six benchmark masses without changing the ATLAS-selected background, likelihood windows, detector response, or nuisance model. Table~\ref{tab:selbkgdiag} compares these roots with the headline absolute-normalization values.

\begin{table*}[tbp]
\centering
\caption{Normalization robustness diagnostic. The second column is the headline absolute-normalization reach. The third column applies the reconstructed fractional deformation to the total selected-background template, $s_i^{\rm sel}=r_i b_i$. The last column gives the relative shift. Because $b_i$ includes non-$t\bar t$ backgrounds, the third column is a diagnostic and not the category-resolved Eq.~\eqref{eq:rationorm} result.}
\label{tab:selbkgdiag}
\begin{tabular}{rccc}
\toprule
$\mph$ [GeV] & absolute [TeV] & selected-background diagnostic [TeV] & shift \\
\midrule
400  & 1.9358 & 2.7064 & $+39.8\%$ \\
600  & 2.4886 & 2.7264 & $+9.6\%$ \\
800  & 2.8235 & 2.6189 & $-7.2\%$ \\
1000 & 3.0030 & 2.4223 & $-19.3\%$ \\
1200 & 2.5616 & 1.9966 & $-22.1\%$ \\
1500 & 2.1854 & 1.4701 & $-32.7\%$ \\
\bottomrule
\end{tabular}
\end{table*}

All six diagnostic roots are monotonic crossings and close numerically at $\CLs=0.05$. The pronounced maximum of the absolute-normalization curve near 1 TeV is not preserved: the selected-background diagnostic is nearly flat at 400--600 GeV and then decreases with mass. This confirms directly that the precise location and height of the 1 TeV maximum are normalization-prescription dependent. The behavior is also analytically transparent. In the present hybrid construction Eq.~\eqref{eq:selbkgdiag} is equivalent to $s_i^{\rm sel}=r_B s_i^{\rm abs}$ at fixed mass. Since the reach is interference dominated and the leading deformation scales approximately as $s\propto\Lr^{-2}$, one expects schematically $\Lambda_{r,{\rm sel}}^{95}\simeq\Lambda_{r,{\rm abs}}^{95}\sqrt{r_B}$. This estimate reproduces the exact reprofiled trend to percent-level accuracy over most of the scan.

The diagnostic therefore does not invalidate the absolute-normalization likelihood; rather, it identifies which feature is not robust under a change of normalization convention. The few-TeV sensitivity over the intermediate-mass region and the interference-driven nature of the signal remain, whereas the detailed mass dependence of the headline reach should not be interpreted as a model-independent prediction. A fully category-resolved implementation of Eq.~\eqref{eq:rationorm} remains preferable once selected $t\bar t$ and non-$t\bar t$ components and a radion-specific residual acceptance correction are available.

\subsection{Interference dominance at 800 GeV}
\begin{figure}[htbp]
\centering
\includegraphics[width=\columnwidth]{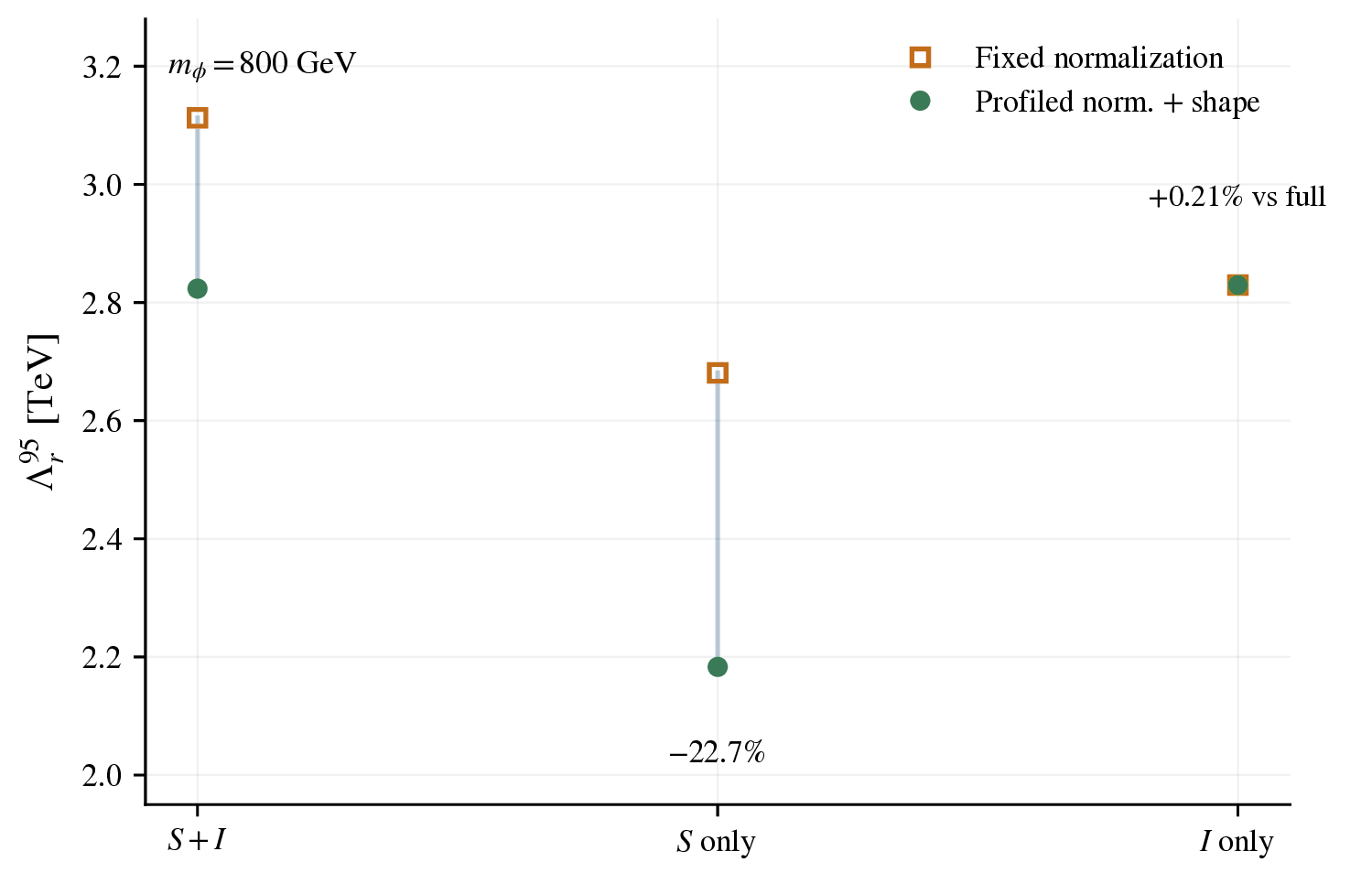}
\caption{Median expected 95\% $\CLs$ reach at $m_\phi=800$ GeV for a diagnostic decomposition of the radion contribution. Open squares show fixed background normalization and filled circles show the result after profiling the normalization and correlated shape nuisance. Removing interference lowers the profiled reach by 22.7\%, whereas the interference-only diagnostic differs from the full coherent prediction by only $+0.2\%$. The $S$-only and $I$-only cases are diagnostic decompositions of the same physical prediction, not independent radion models.}
\label{fig:components}
\end{figure}

The component decomposition in Table~\ref{tab:components} and Fig.~\ref{fig:components} quantifies how strongly the reach depends on interference.

\begin{table}[htbp]
\centering
\caption{Median expected 95\% $\CLs$ component study at $\mph=800$ GeV. The $S$-only and $I$-only entries are diagnostic decompositions of the same physical model, not independent radion hypotheses.}
\label{tab:components}
\resizebox{\columnwidth}{!}{%
\begin{tabular}{lccc}
\toprule
component & fixed [TeV] & profiled [TeV] & shift vs full \\
\midrule
$S+I$ & 3.11 & 2.82 & 0 \\
$S$ only & 2.68 & 2.18 & $-22.7\%$ \\
$I$ only & 2.83 & 2.83 & $+0.2\%$ \\
\bottomrule
\end{tabular}}
\end{table}

The physical interpretation is that the resonance-squared contribution is small after the radion-specific reweighting, while the sign-changing interference morphology is difficult for a floating background normalization to absorb. The sensitivity is therefore shape driven rather than bump driven.

\section{Robustness and the threshold-sensitive 400 GeV point}
The signal-efficiency envelope is deliberately asymmetric because it is constructed from three public ATLAS signal models rather than from a symmetric nuisance prescription. Relative to the nominal profiled reach, the low-efficiency edge changes the reach by $-0.16\%$ to $-5.40\%$ across the mass grid, while the high-efficiency edge changes it by $+2.81\%$ to $+21.78\%$. These variations are therefore interpreted as a signal-model envelope, not as an official ATLAS systematic uncertainty.

Replacing the nominal low-mass response with a common 10\% reconstructed-$\mtt$ resolution raises the reach by 12.60\% at 400 GeV, 8.62\% at 600 GeV, and 3.54\% at 800 GeV. The 1000--1500 GeV points are unchanged because their nominal resolution is already 10\%. In this test the continuum template is re-smeared together with the signal and then re-anchored to the same absolute ATLAS selected-background yield, so the variation probes resolution-induced shape migration rather than an artificial normalization change.

\begin{figure}[htbp]
\centering
\includegraphics[width=0.90\columnwidth]{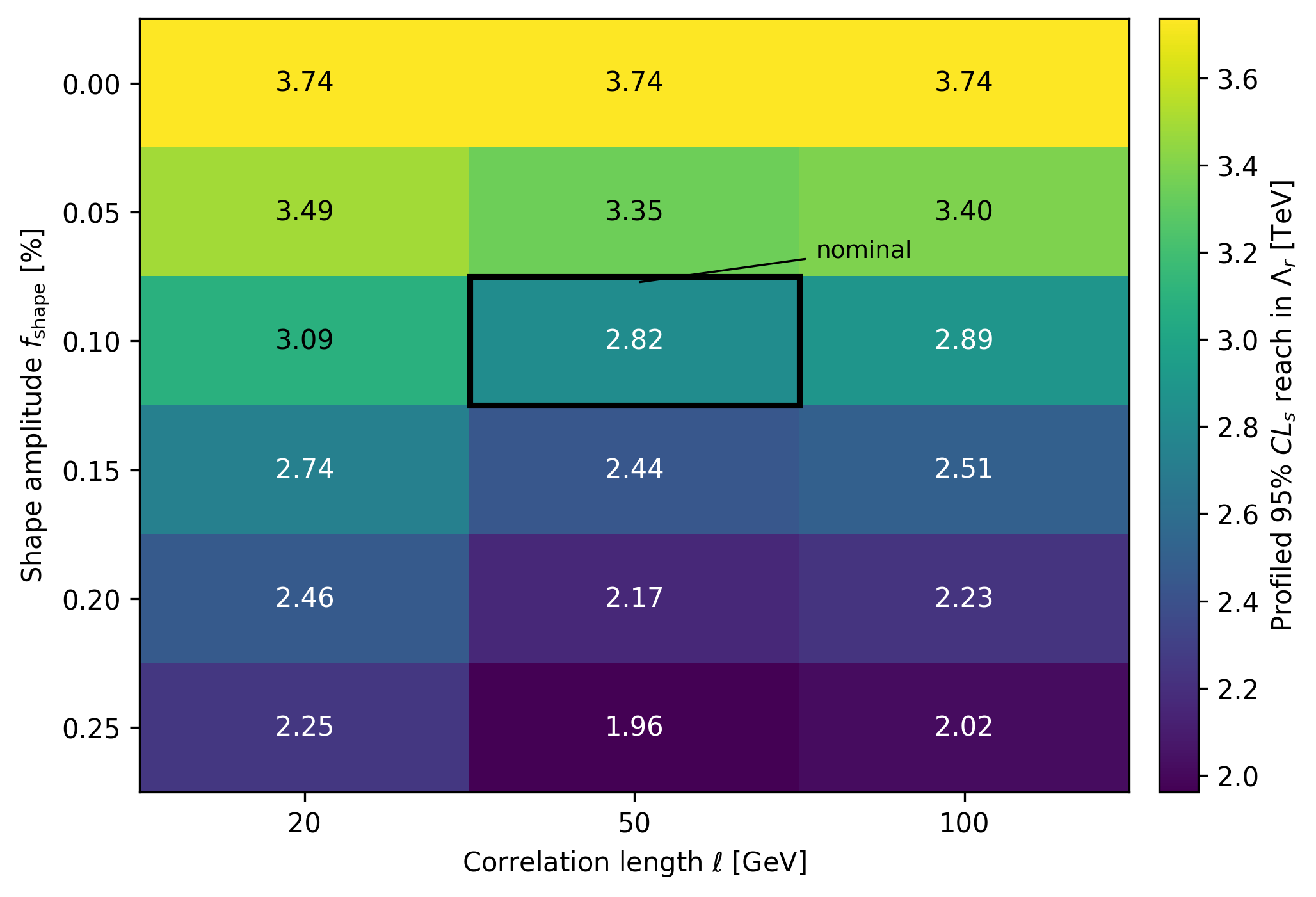}
\caption{Profiled median 95\% $CL_s$ reach in $\Lambda_r$ at
    $m_\phi=800$~GeV as a function of the correlated background-shape
    amplitude $f_{\rm shape}$ and correlation length $\ell$, evaluated
    with the exact-Poisson likelihood. The boxed cell denotes the
    nominal benchmark $f_{\rm shape}=0.1\%$, $\ell=50$~GeV.
    For non-zero shape uncertainty, the intermediate correlation scale
    produces the largest degradation in sensitivity, reflecting its
    stronger overlap with the reconstructed interference deformation.}
\label{fig:fshapeell}
\end{figure}

\begin{figure}[htbp]
\centering
\includegraphics[width=\columnwidth]{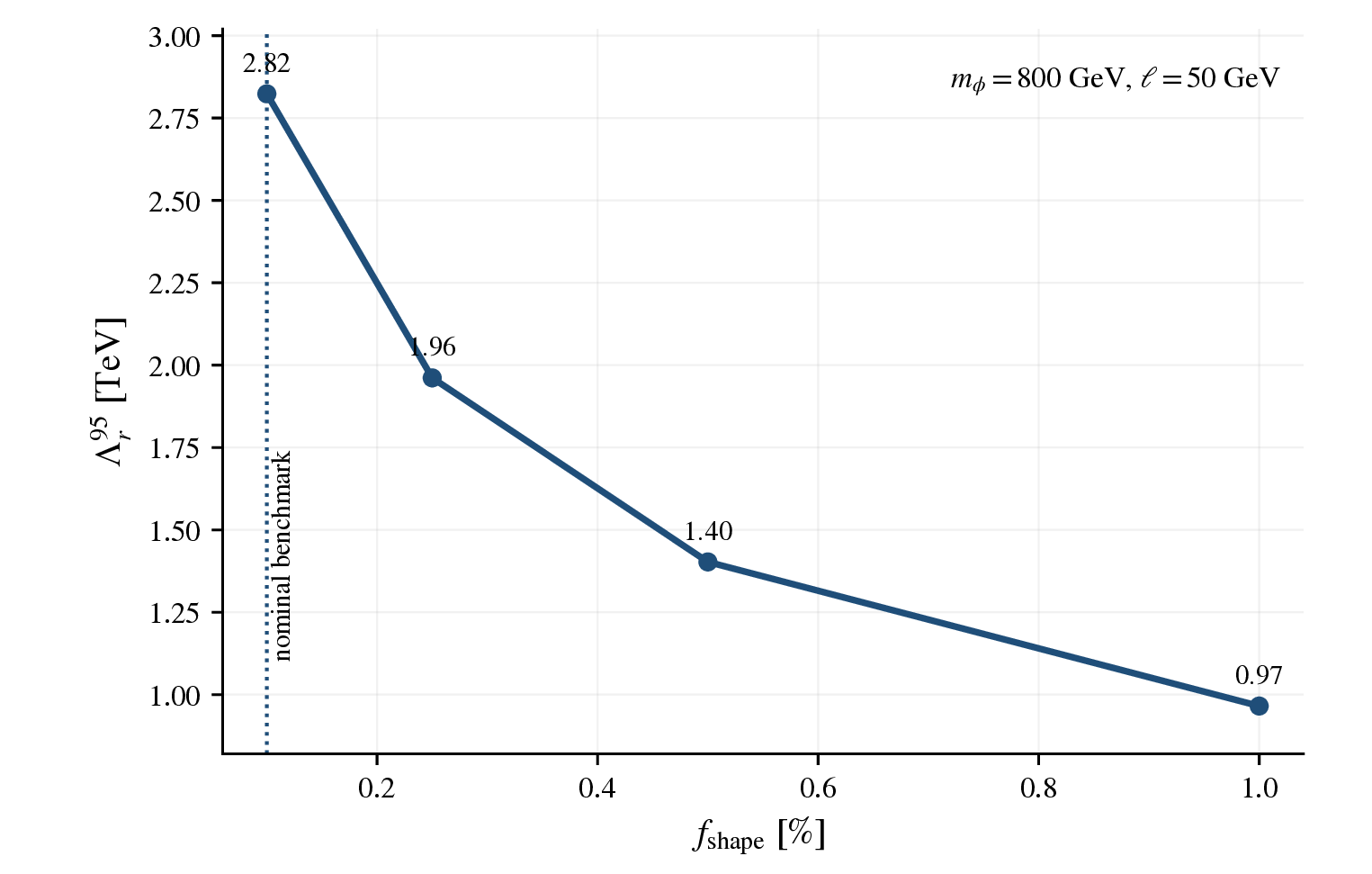}
\caption{Exact-Poisson $\CLs$ robustness of the $m_\phi=800$ GeV profiled reach to the correlated background-shape amplitude at fixed correlation length $\ell=50$ GeV. The reach decreases from 2.82 TeV at the nominal $f_{\rm shape}=0.1\%$ benchmark to 0.97 TeV at 1\%.}
\label{fig:fshape}
\end{figure}

The assumed residual correlated shape uncertainty has a larger impact.
To test both its amplitude and correlation scale within the same
statistical prescription as the headline limits, the exact-Poisson
$CL_s$ profiler is rerun at $m_\phi=800$~GeV over
$f_{\rm shape}=0$--$0.25\%$ and correlation lengths
$\ell=20$, $50$, and $100$~GeV. The resulting profiled median
95\% $CL_s$ reaches are shown in Fig.~\ref{fig:fshapeell}.
The nominal benchmark,
$f_{\rm shape}=0.1\%$ and $\ell=50$~GeV, reproduces the headline
reach exactly,
\[
\Lambda_r^{95}=2.823487~{\rm TeV}.
\]
For every non-zero shape amplitude considered, the intermediate
correlation length $\ell=50$~GeV gives the largest degradation of the
reach, whereas both $\ell=20$ and $100$~GeV retain more sensitivity.
For example, at $f_{\rm shape}=0.1\%$ the corresponding reaches are
$3.086$, $2.823$, and $2.889$~TeV for
$\ell=20$, $50$, and $100$~GeV, respectively. This non-monotonic
dependence on $\ell$ shows that the nuisance impact is controlled by
the overlap of the correlated background-deformation modes with the
signed reconstructed radion template, rather than by a simple rule in
which longer-range correlations are always more damaging.
For the exact-Poisson $\CLs$ likelihood at fixed $\ell=50$ GeV, increasing $\fshape$ from the nominal 0.1\% to 0.25\%, 0.5\%, and 1.0\% changes the profiled reach from 2.82 TeV to 1.96, 1.40, and 0.97 TeV, respectively. Figure~\ref{fig:fshape} makes this dependence explicit; the nominal 0.1\% point should therefore be read as an optimistic shape-control benchmark rather than as experimentally established HL-LHC performance. The large-correlation-length limit is fixed analytically: as $\ell\to\infty$, $K^{\rm shape}_{ij}\to f_{\rm shape}^2 b_i b_j$, a rank-one deformation exactly parallel to the freely profiled normalization direction. The additional shape nuisance therefore becomes redundant with $\alpha$, so its independent effect vanishes in the $\ell\to\infty$ limit. This explains why the most damaging correlation scale can occur at an intermediate $\ell$ comparable to the reconstructed deformation width rather than at arbitrarily large $\ell$.

\subsection{Response-tail and mass-dependent-efficiency diagnostics}
A single Gaussian response is a controlled baseline, but non-Gaussian tails can rotate a sign-changing template relative to the profiled nuisance modes. We therefore do not use the RMS of the deformation as a surrogate for sensitivity. The archived 800 GeV 10\% and 15\% Gaussian responses provide a morphology-level stress test: mixing a 15\% broad component with fractions 10\% and 20\% changes the maximum local deformation by only 0.009 and 0.019 percentage points, respectively, while preserving the peak--dip ordering. These numbers are useful for scale, but they are not converted into a $\Lr$ limit because the underlying selected $B$ and $S+I$ yield vectors are required to rerun the profiler consistently.

The same caution applies to the mass-dependent $\epsilon A$ proxy. The public ATLAS curves are efficiencies for generated resonance benchmarks, not an event-level map $\epsilon A(\mtt)$ for a signed radion deformation. A clipped interpolation through the six anchors changes the local reconstructed ratio by at most 0.038 percentage points in the current proxy study, with the largest effect at 800 GeV. We therefore retain the single-number $\epsilon A(\mph)$ prescription in the quoted likelihood and interpret the proxy only as a morphology diagnostic. A fully quantitative replacement requires either a radion-specific detector sample or direct propagation of the reweighted signed template through the existing profiler.

The 400 GeV point lies only about 55 GeV above $2m_t\simeq345$ GeV. Dedicated LHC measurements demonstrate that this threshold region is sensitive to non-relativistic QCD effects and quasi-bound-state dynamics~\cite{ATLASThreshold2026}; these effects are not included in the present fixed-order radion calculation. For this reason, the primary quantitative interpretation of the paper is restricted to $\mph\ge600$ GeV, while the 400 GeV point is retained as an explicitly auxiliary benchmark.

\section{Experimental context, competing channels, and RS consistency}
ATLAS and CMS have performed interference-aware searches for heavy scalar and pseudoscalar states in the $\ttbar$ channel~\cite{ATLAS2024,CMS2025}. A newer CMS combination also searches broadly for $\ttbar$ resonances in the all-hadronic, single-lepton, and dilepton final states and includes spin-zero interpretations in the 0.5--1.0 TeV region~\cite{CMSB2G2026}. These analyses provide essential context but are not converted directly into $\Lr$ exclusions here. A generic scalar with a top-loop-only production amplitude does not share the radion's anomaly-plus-loop coefficient or its mass-dependent complex phase, and a direct coupling rescaling would not preserve the physical line shape.

\begin{figure}[htbp]
\centering
\includegraphics[width=\columnwidth]{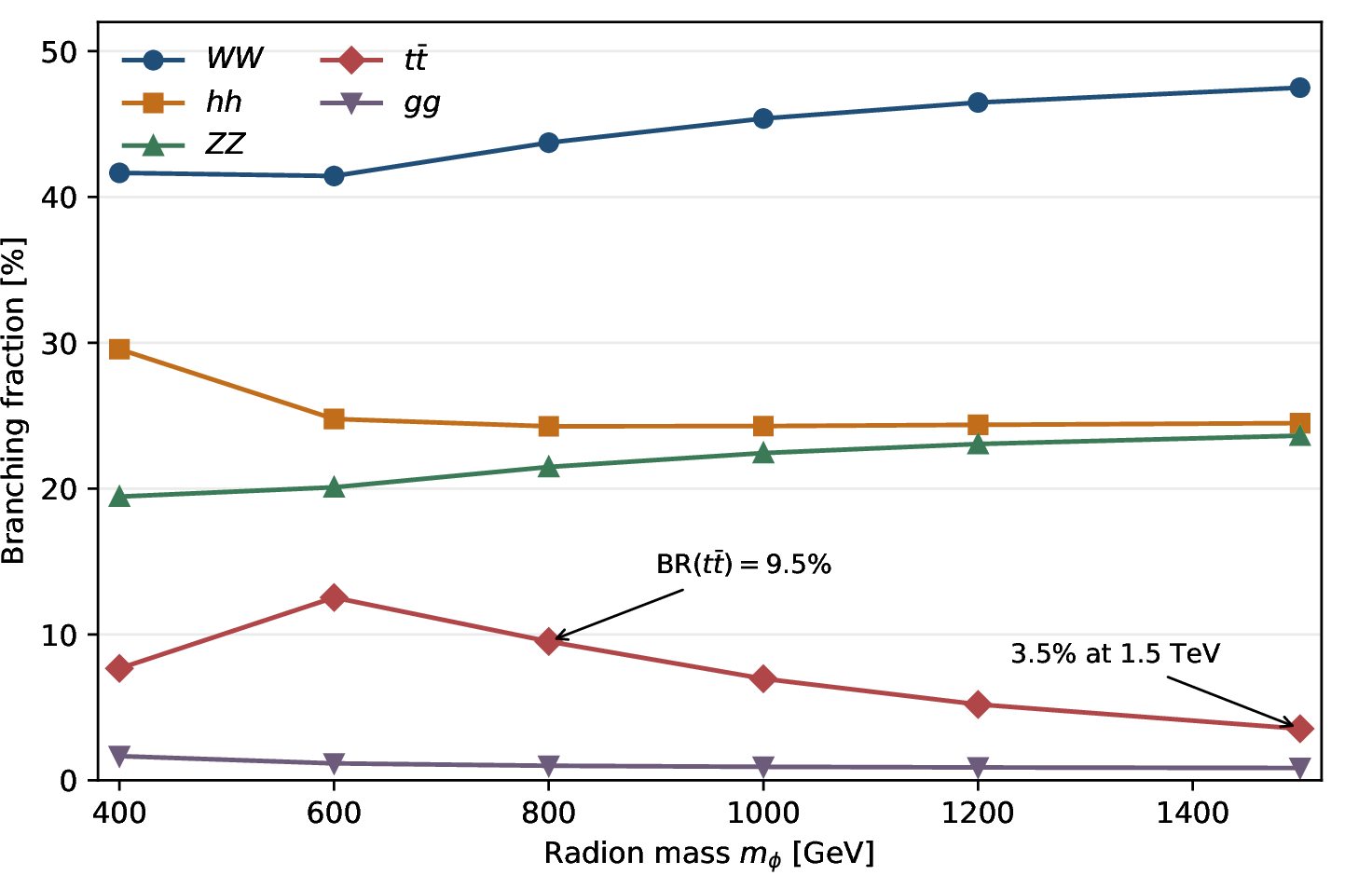}
\caption{Mass dependence of the pure-radion branching fractions used in the analysis. The relative importance of $t\bar t$ decreases above the intermediate-mass region because the bosonic widths grow as $m_\phi^3$ while the fermionic width scales only as $m_\phi m_t^2$.}
\label{fig:branching}
\end{figure}

The $\ttbar$ mode is also not the rate-dominant decay of the pure radion. The same width model used in the likelihood gives the branching fractions in Table~\ref{tab:branching} and Fig.~\ref{fig:branching}. The high-mass behavior is simple: $\Gamma_{t\bar t}\propto m_\phi m_t^2$ up to threshold factors, whereas $\Gamma_{WW}$, $\Gamma_{ZZ}$, and $\Gamma_{hh}$ scale approximately as $m_\phi^3$. Consequently $\mathrm{BR}(t\bar t)$ is 7.7\% at 400 GeV, rises to 12.5\% at 600 GeV as the top threshold opens, is 9.5\% at 800 GeV, and falls to 3.5\% by 1.5 TeV. The $\ttbar$ channel therefore becomes relatively less competitive in rate toward the high-mass end, precisely where the projected interference reach is already declining.

\begin{table}[htbp]
\centering
\caption{Pure-radion branching fractions at $\Lr=3$ TeV from the same partial-width model used in the scan. Percentages are independent of $\Lr$ at fixed mass in the pure-radion limit because all listed partial widths scale as $\Lr^{-2}$.}
\label{tab:branching}
\resizebox{\columnwidth}{!}{%
\begin{tabular}{llllll}
\toprule
$m_\phi$ [GeV] & $WW$ & $hh$ & $ZZ$ & $t\bar t$ & $gg$ \\
\midrule
400  & 41.65\% & 29.56\% & 19.45\% & 7.67\% & 1.66\% \\
600  & 41.44\% & 24.78\% & 20.09\% & 12.53\% & 1.16\% \\
800  & 43.73\% & 24.27\% & 21.49\% & 9.52\% & 1.00\% \\
1000 & 45.38\% & 24.29\% & 22.44\% & 6.96\% & 0.92\% \\
1200 & 46.48\% & 24.38\% & 23.06\% & 5.20\% & 0.88\% \\
1500 & 47.50\% & 24.49\% & 23.63\% & 3.53\% & 0.84\% \\
\bottomrule
\end{tabular}}
\end{table}

Diboson and Higgs-pair searches are therefore intrinsically competitive and can be stronger in parts of the parameter space. ATLAS has combined resonant $hh$ searches over a broad mass range~\cite{ATLASHH2024} and has searched for heavy diboson resonances~\cite{ATLASDiboson2020}; CMS has interpreted boosted $hh\to b\bar b b\bar b$ searches directly in warped-model radion benchmarks~\cite{CMSHHradion2025} and has recently reported a heavy-scalar $ZZ\to4\ell$ search up to 3 TeV~\cite{CMSZZ2026}. We do not overlay those exclusions as one-to-one bounds on the present $\Lr$ convention because their production, width, mixing, and acceptance assumptions differ. The value of the present $\ttbar$ analysis is instead the coherent QCD interference and the finite-top absorptive phase, which are not captured by a positive-resonance rate comparison.

As an internal consistency diagnostic, the first KK-graviton mass may be written
\begin{equation}
\begin{aligned}
m_{G_1}&=x_1ke^{-kL}=\frac{x_1}{\sqrt6}c\Lr,\\
c&=\frac{k}{\bar M_{\rm Pl}},\qquad x_1=3.83170597.
\end{aligned}
\end{equation}
Requiring $m_{G_1}>\mph$ gives
\begin{equation}
\boxed{c_{\min}=\frac{\sqrt6}{x_1}\frac{\mph}{\Lr}.}
\end{equation}
Using $\bar M_{\rm Pl}^2=M_5^3/k$, one also has $k/M_5=c^{2/3}$. Evaluated at the quoted profiled reach, $c_{\min}$ increases from 0.132 at 400 GeV (0.154 at 600 GeV) to 0.439 at 1500 GeV, while the corresponding $(k/M_5)_{\min}$ values span 0.259--0.577. This is not an experimental exclusion; it is an illustrative KK-separation and effective-description consistency diagnostic.

\section{Discussion and limitations}
The amplitude-level result is considerably more robust than the absolute sensitivity number, and the claim boundary is therefore explicit. First, the analysis is the pure-radion limit $\xi=0$, not a Higgs--radion mixing scan. Second, the hard-scattering calculation is leading order in QCD; the finite-top loop is exact at this order, but no NLO signal--background interference calculation is included. Third, the detector treatment is an ATLAS-anchored phenomenological response rather than a full detector simulation, and the public resonance efficiencies are not a radion-specific event-level transfer matrix. Fourth, $f_{\rm shape}=0.1\%$ is an optimistic benchmark for residual shape control, not a measured HL-LHC uncertainty. Fifth, the 400 GeV point is threshold sensitive and is not used for the primary interpretation.

A sixth limitation concerns normalization. The quoted reach table is obtained from the fully reprofiled absolute signal normalization in Eq.~\eqref{eq:signalabs} together with the ATLAS-selected background anchor. The selected-yield deformation formulation in Eq.~\eqref{eq:rationorm} is theoretically attractive because common QCD normalization, PDF, and much of the selection efficiency cancel, but a clean implementation requires the selected $t\bar t$ component separately from non-$t\bar t$ backgrounds and a residual angular-acceptance correction. We have therefore reprofiled the simplified total-selected-background diagnostic of Eq.~\eqref{eq:selbkgdiag} rather than silently identifying it with Eq.~\eqref{eq:rationorm}. The resulting 600--1500 GeV reaches are 2.73, 2.62, 2.42, 2.00, and 1.47 TeV and demonstrate that the pronounced 1 TeV maximum in the absolute-normalization scan is not robust to this change of convention. These values are retained as a normalization stress test, not as replacement headline limits. A fully category-resolved likelihood remains an important extension because it would quantify the residual angular-acceptance and partonic-composition effects directly.

The phase-basis signal and interference samples used in this revision are generated directly at 14 TeV. The selected-background anchor is still transferred from measured 13 TeV ATLAS yields using the inclusive Top++ ratio of Eq.~\eqref{eq:bkgtransfer}; a window-differential transfer would improve the high-mass treatment. The response-tail and mass-dependent-efficiency studies are morphology stress tests only, because a likelihood statement requires the underlying selected yield vectors rather than an RMS proxy. Within these boundaries, the complex radion coefficient closes against a native loop-induced implementation at multiple masses, the common-event phase basis obeys over-constrained identities, and every quoted $95\%$ $\CLs$ root closes at $\CLs=0.05$.

\section{Conclusions}
We have constructed a pure-radion $\ttbar$ interference analysis in which the QCD trace anomaly and the exact finite-top-mass loop are combined at amplitude level. The validated production coefficient
\begin{equation}
C_\phi(\shat)=\frac{2m_t^2}{\Lr^2}\left[7+\frac12A_{1/2}(\shat)\right]
\end{equation}
retains both the correct normalization and the absorptive phase above the top threshold. Equation~\eqref{eq:peakdipsign} fixes the peak--dip ordering analytically: the dispersive contribution is positive below the pole and negative above it in the present convention, while the absorptive component moves the interference zero only about 0.12 GeV below the 800 GeV pole and makes the on-pole interference non-zero.

The fully reprofiled absolute-normalization analysis gives median expected 95\% $\CLs$ reaches of 2.49, 2.82, 3.00, 2.56, and 2.19 TeV for $m_\phi=600,800,1000,1200,$ and 1500 GeV, with the 400 GeV point retained only as a threshold-sensitive auxiliary benchmark. At 800 GeV, removing interference lowers the profiled reach from 2.82 TeV to 2.18 TeV, whereas the interference-only diagnostic gives 2.83 TeV. This interference dominance is the central physics result and is independent of interpreting the radion as a conventional bump.

Several additional diagnostics sharpen the interpretation. The exact-Poisson two-dimensional $(f_{\rm shape},\ell)$ scan shows that the nuisance impact is controlled by template--eigenmode overlap rather than by a monotonic ordering in correlation length, while the extended scan at fixed $\ell=50$~GeV quantifies the strong dependence on the shape amplitude. The pure-radion branching fractions show that $t\bar t$ is not the rate-dominant channel and falls to only 3.5\% at 1.5 TeV, so diboson and Higgs-pair searches can be more competitive in rate; the distinctive role of $t\bar t$ is instead its coherent interference with QCD. The broad-response and mass-dependent-efficiency tests are kept as morphology diagnostics rather than translated into limits through an RMS proxy.

A complementary selected-yield deformation normalization, $s_i=r_i b_i^{t\bar t}$, can reduce sensitivity to common QCD normalization, PDFs, and selection efficiency. As a direct stress test with the available public inputs, we reran the full exact profiler using the total selected background, $s_i^{\rm sel}=r_i b_i$. The corresponding reaches are 2.71, 2.73, 2.62, 2.42, 2.00, and 1.47 TeV for $m_\phi=400,600,800,1000,1200,$ and 1500 GeV, respectively. The 1 TeV maximum is removed, demonstrating that the detailed mass dependence of the absolute-normalization reach is convention dependent. Because the total selected background contains non-$t\bar t$ contributions and no residual scalar-versus-continuum acceptance correction is applied, these values are presented only as a normalization diagnostic rather than as the category-resolved result of Eq.~\eqref{eq:rationorm}. The separate component study at 800~GeV nevertheless shows that the headline sensitivity is strongly interference dominated. A mixed Higgs--radion scan, higher-order QCD interference, and a dedicated threshold-resummed treatment of the 400 GeV region remain natural extensions.

\appendix
\section{Pure-radion partial widths}
\label{app:purewidths}
For completeness, this appendix records the partial widths used to construct the pure-radion total width in the numerical analysis. These expressions correspond to the unmixed limit $\xi=0$ of the trace-coupled radion and follow the standard radion convention~\cite{Bae2000,Dominici2003}. Define
\begin{equation}
 x_X=\frac{m_X^2}{\mph^2},\qquad
 \beta_X=\sqrt{1-4x_X}.
\end{equation}
For a fermion $f$ with colour multiplicity $N_c$, the tree-level width is
\begin{equation}
 \boxed{\Gamma(\phi\to f\bar f)=
 N_c\,\frac{\mph m_f^2}{8\pi\Lr^2}\,\beta_f^3.}
 \label{eq:widthff}
\end{equation}
In particular, $N_c=3$ gives the $t\bar t$ width used in the benchmark scan.

For massive electroweak gauge bosons,
\begin{align}
 \boxed{\Gamma(\phi\to W^+W^-)}&=
 \frac{\mph^3}{16\pi\Lr^2}\,
 \beta_W\left(1-4x_W+12x_W^2\right),
 \label{eq:widthww}\\
 \boxed{\Gamma(\phi\to ZZ)}&=
 \frac{\mph^3}{32\pi\Lr^2}\,
 \beta_Z\left(1-4x_Z+12x_Z^2\right).
 \label{eq:widthzz}
\end{align}
The factor of two between the $WW$ and $ZZ$ prefactors accounts for the identical-particle factor in the neutral final state.

For Higgs pairs, the pure-radion trace coupling gives
\begin{equation}
 \boxed{\Gamma(\phi\to hh)=
 \frac{\mph^3}{32\pi\Lr^2}\,
 \beta_h\left(1+2x_h\right)^2.}
 \label{eq:widthhh}
\end{equation}
The gluonic width is evaluated with the same finite-top-mass form factor used in the interference calculation,
\begin{equation}
 F_\phi(\shat)=7+\frac12 A_{1/2}(\shat),
\end{equation}
so that
\begin{equation}
 \boxed{\Gamma(\phi\to gg)=
 \frac{\as^2(\mph)\,\mph^3}{32\pi^3\Lr^2}
 \left|F_\phi(\mph^2)\right|^2.}
 \label{eq:widthgg}
\end{equation}
Thus the width model used in the scan is
\begin{equation}
 \boxed{\Gamma_\phi=
 \Gamma_{WW}+\Gamma_{ZZ}+\Gamma_{t\bar t}+\Gamma_{hh}+\Gamma_{gg}.}
 \label{eq:widthtotal}
\end{equation}
All five terms are proportional to $\Lr^{-2}$, yielding the rescaling in Eq.~\eqref{eq:widthscaling}.

As a numerical closure at the reference point $\mph=800$ GeV and $\Lr=3$ TeV, using $m_t=172.5$ GeV and $\as(800\,\mathrm{GeV})\simeq0.089585$, the implementation gives
\begin{align}
 \Gamma_{WW}&=1.06527~\mathrm{GeV}, &
 \Gamma_{ZZ}&=0.52347~\mathrm{GeV},\\
 \Gamma_{t\bar t}&=0.23188~\mathrm{GeV}, &
 \Gamma_{hh}&=0.59136~\mathrm{GeV},\\
 \Gamma_{gg}&=0.02426~\mathrm{GeV}.&&
\end{align}
Their sum is
\begin{equation}
 \Gamma_\phi^{\rm calc}(800~\mathrm{GeV})=2.43623~\mathrm{GeV}\simeq2.44~\mathrm{GeV},
\end{equation}
which is the width used for the 800 GeV event benchmark up to the quoted rounding. The remaining mass points are generated with the same expressions and inputs, with $\as$ evaluated at the corresponding radion mass.

\section{Validation details}
At $\mph=800$ GeV and $\Lr=3$ TeV, the exact Poisson profile with the correlated shape nuisance gives
\begin{equation}
\sqrt{q_A}=1.74093590,
\end{equation}
compared with the Gaussian-covariance diagnostic $Z=1.74080983$, a relative difference of 0.0072\%. At the final profiled 95\% root, $\sqrt{q_A}=1.9599639846$ and $\CLs=0.0500000000$. The corresponding fitted normalization is $\alpha=0.9991963$, with $\lVert z\rVert=1.5515$ and $\max|z_i|=0.4347$. The final roots are monotonic in $\Lr$, providing a direct diagnostic against root-finding or profiling pathologies.

\section{Resolution-smearing validation}
\begin{figure}[htbp]
\centering
\includegraphics[width=\columnwidth]{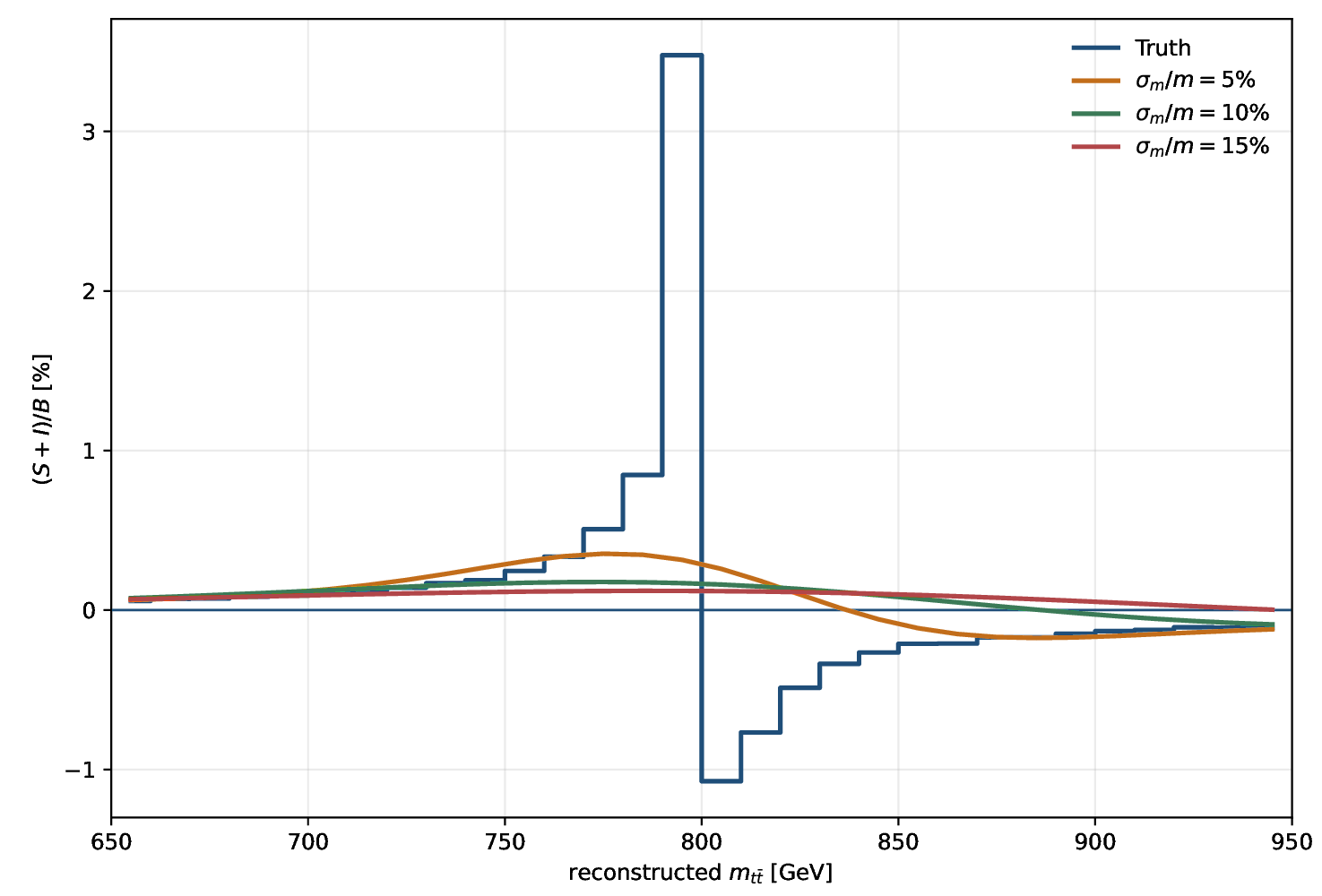}
\caption{Resolution-smearing study for the 800 GeV benchmark. The narrow truth-level interference structure is strongly diluted at the 10\% reconstructed-mass scale.}
\label{fig:smearing}
\end{figure}

The response model was separately tested by smearing the same 800 GeV physical template with several Gaussian resolutions. As the reconstructed resolution degrades, the narrow truth-level peak--dip is progressively washed out, confirming that detector resolution controls the observable morphology in the narrow-radion regime; the comparison is shown explicitly in Fig.~\ref{fig:smearing}.

\section{Illustrative KK-separation diagnostic}
\begin{figure}[htbp]
\centering
\includegraphics[width=\columnwidth]{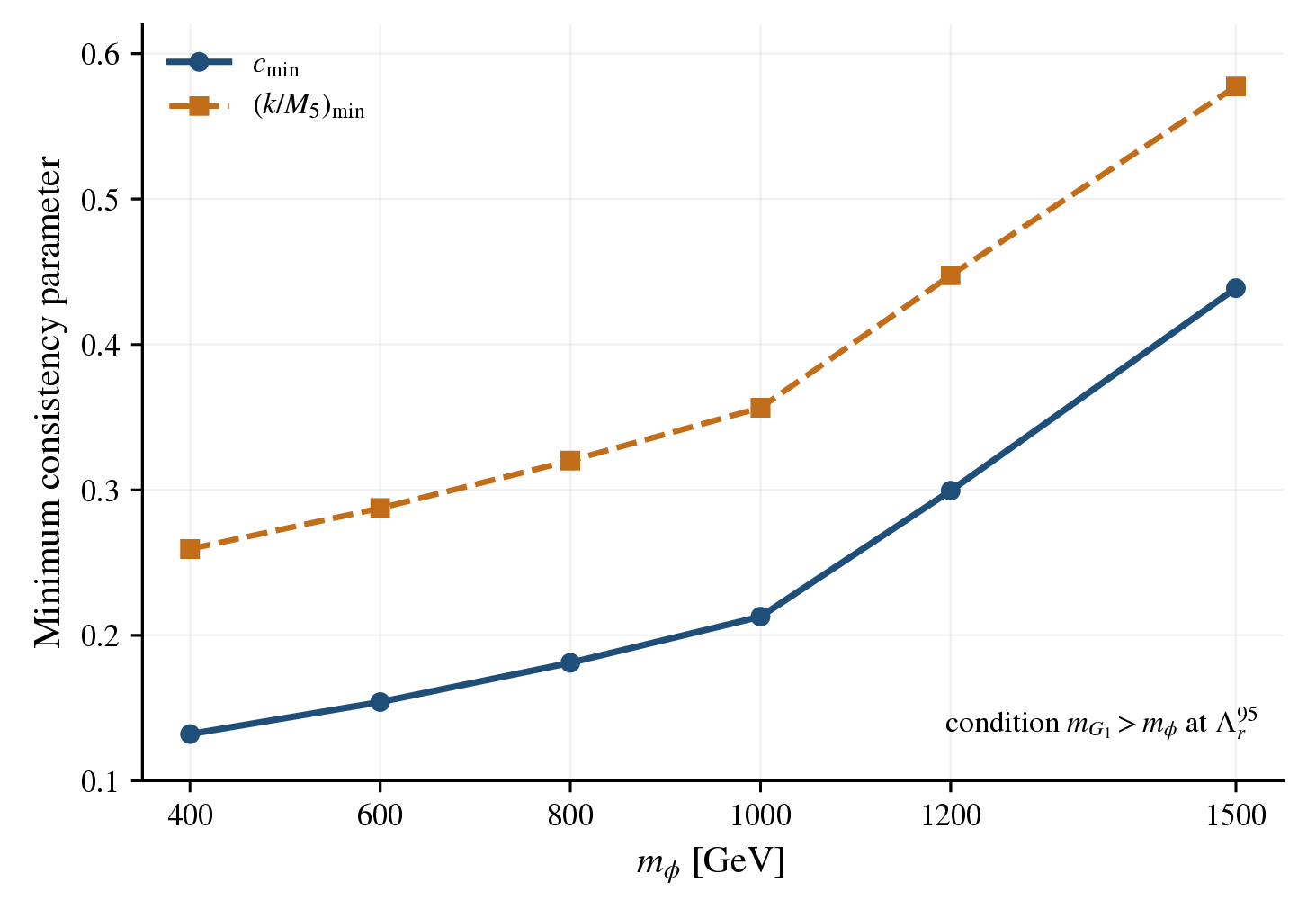}
\caption{Illustrative KK-separation condition $m_{G_1}>\mph$ evaluated at the final profiled 95\% $\CLs$ reach. The plotted $c_{\min}$ and $(k/M_5)_{\min}$ values are model-consistency diagnostics, not experimental exclusions.}
\label{fig:uv}
\end{figure}

For reference, the RS consistency quantity $c_{\min}$ and the corresponding $(k/M_5)_{\min}$ evaluated at the final $\CLs$ reach are shown in Fig.~\ref{fig:uv}. The purpose of this figure is only to visualize the loss of spectral separation at high radion mass.

\section*{Statements and Declarations}
\paragraph{acknowledgments:}
The work of A.~Bellagroudi is funded by the National Center for Scientific and Technical Research (CNRST) under the PhD-Associate Scholarship (PASS). 
\paragraph{Data availability:}
The experimental inputs used for the detector response and selected-background normalization are public and are cited through the corresponding ATLAS publications and HEPData record. The processed radion templates, numerical likelihood inputs, and custom analysis scripts underlying the results are available from the authors upon reasonable request. Numerical benchmark values needed to reproduce the quoted results are documented in the text and appendices.
\balance

\end{document}